\documentclass[10pt,aps,prb,showpacs,twocolumn,amsmath,amssymb,superscriptaddress]{revtex4-1}

\usepackage{bbm}
\usepackage{mathtools}
\usepackage{graphicx}
\usepackage{appendix}
\usepackage{epstopdf}
\usepackage{multirow}
\usepackage{color}
\DeclareMathOperator{\str}{str}
\DeclareMathOperator{\Str}{Str}
\DeclareMathOperator{\sdet}{sdet}
\DeclareMathOperator{\tr}{tr}
\DeclareMathOperator{\Tr}{Tr}

\DeclareMathOperator{\arctanh}{arctanh}

\DeclareMathOperator{\im}{Im}
\DeclareMathOperator{\re}{Re}
\DeclareMathOperator{\var}{var}
\DeclareMathOperator{\diag}{diag}
\DeclareMathOperator{\Ai}{Ai}
\DeclareMathOperator{\Bi}{Bi}
\newcommand{\Z}{\mathbb{Z}}

\allowdisplaybreaks

\begin{document}

\title{Semiclassical electron transport at the edge of a two-dimensional topological insulator: Interplay of protected and unprotected modes}

\author{E.\ Khalaf}
\affiliation{Max Planck Institute for Solid State Research, Heisenbergstr.\ 1, 70569 Stuttgart, Germany}
\author{M.\ A.\ Skvortsov}
\affiliation{Skolkovo Institute of Science and Technology, 143025 Skolkovo, Russia}
\affiliation{L.\ D.\ Landau Institute for Theoretical Physics RAS, 142432 Chernogolovka, Russia}
\affiliation{Moscow Institute of Physics and Technology, 141700 Dolgoprudny, Russia}
\author{P.\ M.\ Ostrovsky}
\affiliation{Max Planck Institute for Solid State Research, Heisenbergstr.\ 1, 70569 Stuttgart, Germany}
\affiliation{L.\ D.\ Landau Institute for Theoretical Physics RAS, 142432 Chernogolovka, Russia}

\begin{abstract}
We study electron transport at the edge of a generic disordered two-dimensional topological insulator, where some channels are topologically protected from 
backscattering. Assuming the total number of channels is large, we consider the edge as a quasi-one-dimensional quantum wire and describe it in terms of a 
non-linear sigma model with a topological term. Neglecting localization effects, we calculate the average distribution function of transmission probabilities as 
a function of the sample length. We mainly focus on the two experimentally relevant cases: a junction between two quantum Hall (QH) states with different 
filling factors (unitary class) and a relatively thick quantum well exhibiting quantum spin Hall (QSH) effect (symplectic class). In a QH sample, the presence 
of topologically protected modes leads to a strong suppression of diffusion in the other channels already at the scales much shorter than the localization 
length. On the semiclassical level, this is accompanied by the formation of a gap in the spectrum of transmission probabilities close to unit transmission, 
thereby suppressing shot noise and conductance fluctuations. In the case of a QSH system, there is at most one topologically protected edge channel leading to 
weaker transport effects. In order to describe `topological' suppression of nearly perfect transparencies, we develop an exact mapping of the semiclassical 
limit of the one-dimensional sigma model onto a zero-dimensional sigma model of a different symmetry class, allowing us to identify the distribution of 
transmission probabilities with the average spectral density of a certain random-matrix ensemble. We extend our results to other symmetry classes with 
topologically protected edges in two dimensions.
\end{abstract}


\maketitle

\section{Introduction}

Topological insulators and superconductors are a subject of intense theoretical and experimental studies in the past years \cite{Hasan10, Moore09, Qi11} due to 
their fascinating electronic properties and potential applications ranging from spintronics \cite{Pesin12} to quantum computations. \cite{Nayak08} The 
distinctive feature of these materials is the presence of topologically protected metallic edge or surface modes on the background of a gapped bulk spectrum. 
Historically, the first example of a topological insulator was provided by the two-dimensional (2D) electron gas in a strong magnetic field exhibiting quantum 
Hall effect \cite{Klitzing80, Prange87} (QHE). When the chemical potential is tuned into a gap between Landau levels (quantum Hall plateau regime), electron 
transport is due to chiral one-dimensional modes at the edge of the sample.\cite{Halperin82, Thouless82} These edge channels have a topological origin and evade 
Anderson localization in the presence of disorder thus giving rise to the extremely accurate quantization of Hall conductance. The state of the system is 
characterized by an integer topological invariant \cite{Thouless82}(Chern number) corresponding to the number of edge modes. Hence, QHE gives an example of 
$\mathbb{Z}$ topological insulator.

Another type of 2D topological insulators was discovered in HgTe quantum wells exhibiting the quantum spin-Hall effect (QSHE).\cite{Bernevig06, Koenig07, 
Koenig08} This is an analog of QHE in a system with strong spin-orbit interaction and preserved time-reversal symmetry.\cite{Qi11, Hasan10, Bernevig06} 
Spin-orbit coupling leads to inversion of the band gap and the appearance of the spin-polarized counter propagating edge states. These edge modes are partially 
protected due to Kramers theorem. If disorder scattering preserves time-reversal invariance and the number of edge channels is odd, one edge mode remains immune 
to Anderson localization. Due to the presence of two edges, the longitudinal conductance of the sample takes the quantized value of $2e^2/h$. The distinction 
between ordinary and topological insulator in this case is given by the parity of the number of edge modes. Hence, this type of system is named $\mathbb{Z}_2$ 
topological insulator.

Soon after, topological insulator states were discovered in three-dimensional (3D) Bi alloys (BiAs, BiTe, BiSe, etc.)\ and later in many other related 
compounds.\cite{Hsieh08, Xia09} These materials have gapped bulk spectrum and massless 2D Dirac states at the surface. The surface states are also topologically 
protected from localization in the presence of disorder as long as the time-reversal symmetry is preserved. Following the discovery of 3D topological 
insulators, a complete classification of possible topological phases in systems of free fermions in any spatial dimension and symmetry class was developed. 
\cite{Kitaev09, Ryu10, Schnyder09} Disordered systems are classified according to their symmetries, giving rise to 10 symmetry classes. \cite{Altland97} In 
each spatial dimension, three of these classes may exhibit $\Z$ topological states (with any integer number of protected edge/surface modes) and two classes 
host $\Z_2$ topological states (at most one protected edge/surface mode). Symmetry classification of possible topological states gives rise to the ``periodic 
table'' of topological insulators.\cite{Schnyder09, Kitaev09}

The 2D QHE state belongs to the unitary class (A) with all symmetries broken. In addition, $\Z$ topological states in 2D are also possible in superconductors 
with broken time-reversal symmetry and either preserved (class C) or broken (class D) spin-rotation symmetry. These two cases, although not yet realized 
experimentally, have the names of spin quantum Hall effect (SQHE) and thermal quantum Hall effect (TQHE), respectively.\cite{Senthil99, Read00} The QSHE 
appears in the system with preserved time-reversal symmetry but broken spin symmetry (symplectic class AII). A similar $\Z_2$ topological state is also possible 
in a superconductor with the same symmetries (class DIII).\cite{Volovik09}

In this paper, we study the problem of electron transport at the edge of a 2D topological insulator in the presence of disorder when both topologically 
protected and ordinary (unprotected) states coexist. Whenever the sample length exceeds a certain characteristic length $\xi$, the unprotected channels are 
expected to be localized by disorder. In the opposite limit of short samples, both topologically protected and diffusive channels contribute to transport and 
the question of interplay between them becomes important. In order to study how transport is distributed among different channels, we consider the average 
distribution of transmission probabilities $\rho(T)$. This encodes information on the full counting statistics of the sample including the conductance, shot 
noise, and all higher moments of charge transfer. In the absence of topologically protected channels, the distribution function is universal\cite{Dorokhov84, Nazarov94} in 
the diffusive ($L \ll \xi$) limit.  Our main objective is then to investigate how this universal diffusive behavior is altered in the presence of topological 
protection.

\begin{figure}
\center
\includegraphics[width=0.6\columnwidth]{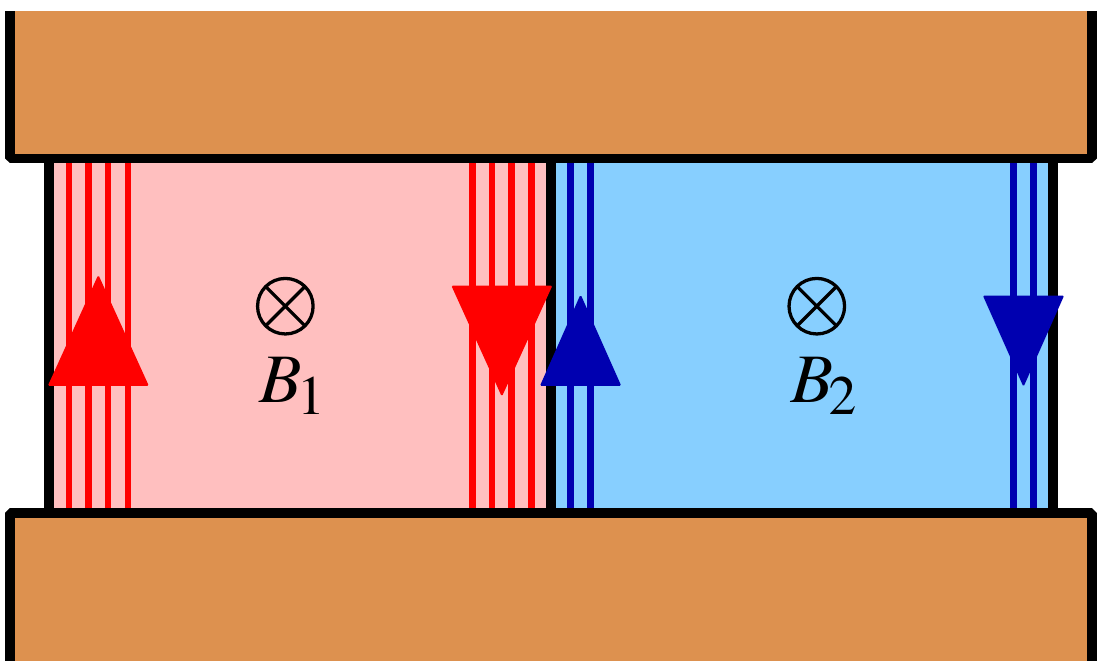}
\caption{Schematic setup of a junction between two quantum Hall systems with different filling factors.}
\label{QH}
\end{figure}

One possible realization of a $\mathbb{Z}$ topological insulator with both protected and diffusive edge states is given by a junction of two quantum Hall 
systems with different filling factors shown schematically in Fig.\ \ref{QH}. In this setup, the boundary between the two parts of the sample hosts a different 
number of right- and left-moving modes, $n_R$ and $n_L$ respectively. This system realizes $\Z$ topological insulator state of the unitary symmetry class A. 
When the two quantum Hall samples are decoupled, all the edge channels conduct perfectly and the overall conductance is\cite{Buttiker85} $G_\text{tot} = N 
e^2/h$ with $N = n_R + n_L$. Coupling at the interface between the two parts of the sample gives rise to backscattering thus suppressing the conductance. The 
backscattering eventually localizes all the channels at the interface except for $m = n_R - n_L$ topologically protected modes. The total conductance is then 
$G_\text{tot} = \max\{n_R, n_L\} e^2/h$. We will separate it into two contributions:
\begin{equation}
 G_\text{tot}
  = \frac{e^2}{h}\, \frac{n_R + n_L}{2} + G.
 \label{Gtotal}
\end{equation}
The first term is due to the outer edges averaged with respect to right-to-left and left-to-right direction of the current. The second term is the conductance
of the middle part averaged in the same manner. Naturally, the total conductance $G_\text{tot}$ is independent of the current direction. In this paper, we will 
consider the transport properties of the middle part only.

A contact between two quantum Hall states with different filling factors was realized in a series of ex\-periments,\cite{Grayson05, Grayson07, Grayson08, 
Steinke08} where a 2D electron gas was confined to the surface of a crystal with an edge between two faces. The filling factors were tuned by changing the 
orientation of the sample in the uniform external magnetic field. Another possible realization of a system with unequal number of counter-propagating chiral 
modes is a Weyl semimetal in a magnetic field which also belongs to the symmetry class A.\cite{Hosur13} Landau levels in such a system consist of states 
confined in the plane perpendicular while having a dispersion along the field direction, with the lowest Landau level having a definite chirality for each Weyl 
point. If disorder does not scatter between Weyl nodes, the system is effectively a quasi-1D wire with a number of chiral topologically protected channels $m$ 
equal to the Landau level degeneracy. Higher (non-chiral) filled Landau levels provide unprotected transport channels carrying current in both directions along 
the magnetic field.\cite{Altland15}

\begin{figure}
\center
\includegraphics[width=0.6 \columnwidth]{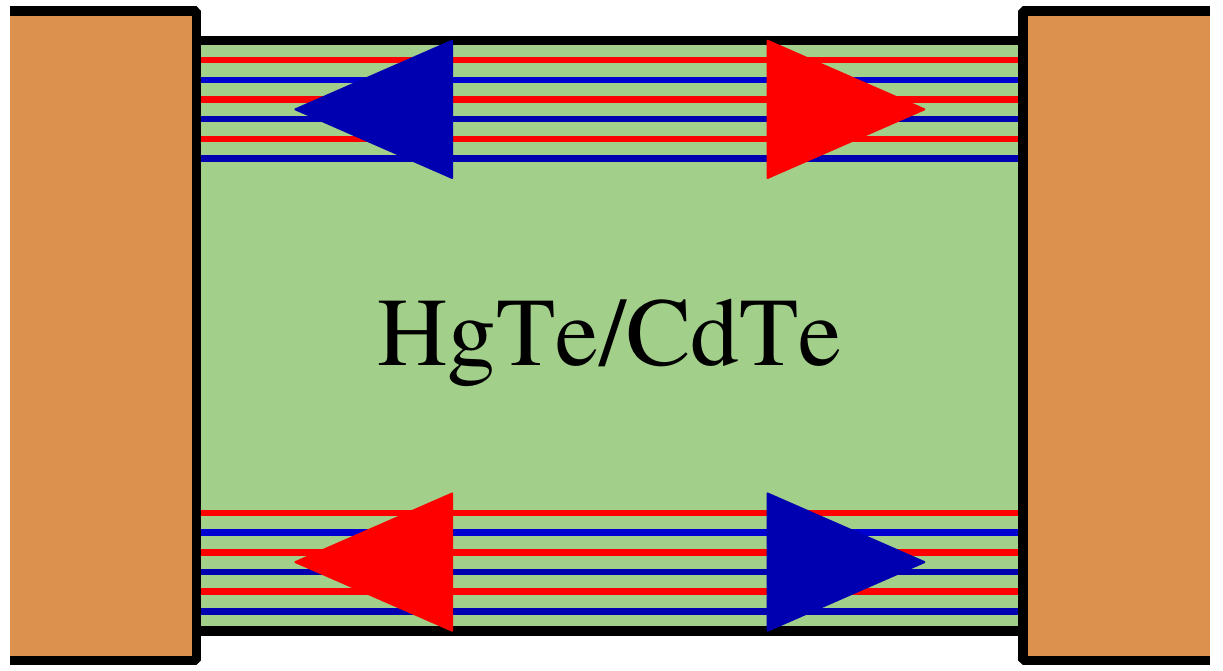}
\caption{Schematic setup of a relatively thick 2D quantum well exhibiting quantum spin Hall effect with a large number of edge channels.}
\label{QSH}
\end{figure}

A possible realization of a $\Z_2$ topological insulator with both protected and unprotected edge modes is given by a relatively thick HgTe/CdTe quantum well, 
see Fig.\ \ref{QSH}. When the width of the HgTe layer exceeds $6.3$\,nm, a pair of counter-propagating topologically protected edge modes 
appears.\cite{Koenig07, Koenig08, Bernevig06} As the width of the quantum well is increased further, additional edge modes become available. Topology protects 
only a single edge channel in the case when their total number is odd. Another realization of the same symmetry is given by a doped metallic carbon nanotube. 
\cite{Ando02, Takane04a, Takane04b} At each of the two valleys (K and K$'$) one protected channel coexists with a number of ordinary channels depending on the 
electron concentration. The role of spin in this case is played by the sublattice index (pseudospin) while the real electron spin remains degenerate. The total 
conductance of the $\mathbb{Z}_2$ topological insulator includes the contribution of two edges, cf.\ Fig.\ \ref{QSH}, or two valleys in the case of the 
nanotube,
\begin{equation}
 G_\text{tot}
  = 2 G.
\end{equation}
We will discuss only the half of this total conductance assuming the disorder does not couple opposite edges/valleys.

On the technical level, the problem of quantum transport is described by a field theory that takes the form of the supersymmetric non-linear sigma 
model.\cite{Efetov99, Mirlin00} The presence of topologically protected channels manifests itself as an additional topological term in the sigma-model action. 
We consider the limit of a short sample, $L \ll \xi$, and analyze the corresponding sigma model in the semiclassical limit. Our main result is that the 
existence of topologically protected channels at the edge of $\mathbb{Z}$ topological insulator leads to a gap in the spectrum of transmission probabilities 
close to unit transmission. 
The appearance of this gap can be understood on a qualitative level in terms of eigenvalue repulsion in a random system where a large number of eigenvalues are pinned at a fixed position. This happens in our system due to the topologically protected channels pinned at unit transmission.
It leads to a reduction in the Fano factor and mesoscopic conductance fluctuations signaling suppressed diffusion of the unprotected channels.  
Quantitatively, the suppression of diffusion due to $m$ protected channels happens at a length scale $\sim \xi/m$ that is much shorter than the localization 
length $\xi$ in the absence of topological protection. In the $\mathbb{Z}_2$ insulator, a similar suppression of diffusion is observed, but is not as 
pronounced. We also show that the average transmission distribution function in the short sample limit can be fully obtained from a zero-dimensional sigma 
model of a different symmetry class.

Throughout the paper, we discuss various transport properties. They represent charge (electrical) transport for systems with electron number conservation
(non-superconducting classes A and AII), spin transport for superconductor with spin rotational symmetry (class C), and thermal transport for superconductors 
without spin rotational symmetry (classes D and DIII). The quantum of conductance in these three cases is \cite{Senthil99, Read00}
\begin{equation}
 G_0
  = \begin{dcases}
      e^2/h, &\text{A, AII}, \\
      2 e^2/h, &\text{C}, \\
      \pi^2 k_B^2 T/3 h, &\text{D, DIII}.
    \end{dcases}
\end{equation}

The paper is organized as follows. In Sec. \ref{formalism}, we review the general theory of electron transport in quasi-1D systems in terms of the matrix
Green function\cite{Nazarov94} and derive the supersymmetric non-linear sigma model with the topological term and source fields. A semiclassical solution of the
sigma model is developed in Sec.\ \ref{QA} for $\mathbb{Z}$ topological insulators. This solution yields the average distribution function of transmission
probabilities. We also derive explicit expressions for the average conductance, Fano factor, and mesoscopic conductance fluctuations. A general mapping of the
semiclassical 1D sigma model onto a random matrix problem is established in Sec.\ \ref{M0D}. This yields a detailed description of the average transmission
distribution in the vicinity of perfect transmission for all symmetry classes. We also discuss the universal crossover at the semiclassical edge of the 
spectrum. The main results are summarized in Sec.\ \ref{Summary}. Some technical details are relegated to two appendixes.

\section{General formalism}
\label{formalism}

\subsection{Transport properties of a quasi-1D system}
\label{MGFF}

We consider the edge of a 2D system (Figs.\ \ref{QH} and \ref{QSH}), which can be thought of as a quasi-1D wire between two perfect metallic leads. Transport 
properties of the system are fully determined by the matrix of transmission amplitudes $t_{mn}$ acting in the space of 1D channels. Each matrix element 
$t_{mn}$ is the probability amplitude for an electron that enters the sample from one lead in the $n$-th channel to be transmitted to another lead in the 
$m$-th channel. While the matrix $t$ depends on the choice of the channel basis in the leads, the eigenvalues of the matrix $t^{\dagger} t$ represent the full 
set $\{T_\alpha\}$ of observable transmission probabilities of the system.

Different transport quantities, such as conductance $G$ and Fano factor $F$, can be expressed in terms of the transmission matrix as\cite{Landauer57, Landauer70, Buttiker85}
\begin{equation}
G = G_0 \tr t^{\dagger} t, \qquad F = 1 - \frac{\tr(t^{\dagger} t)^2}{\tr t^{\dagger} t}.
\end{equation}
The complete distribution of transmission probabilities is encoded in the generating function
\begin{equation}
\mathcal{F}(z) = \sum_{n=1}^{\infty} z^{n-1} \tr (t^{\dagger} t)^n = \tr \left(\frac{t^{\dagger} t}{1 - z t^{\dagger} t} \right).
\end{equation}
This generating function yields all the moments of electron transport, including conductance and noise, as its derivatives in $z$ taken at $z = 0$. Hence,
$\mathcal{F}(z)$ contains information about the full counting statistics of the system.\cite{Levitov93, Lee96}

An equivalent description of transport is given by the distribution function of transmission probabilities
\begin{equation}
 \rho(T) = \tr \delta(T - t^\dagger t).
\end{equation}
This function gives the total number of channels with transmission probability $T$. The two functions $\mathcal{F}(z)$ and $\rho(T)$ are related by the
following identities:
\begin{gather}
 \mathcal{F}(z) = \int_0^1 \frac{\rho(T)\, T\, dT}{1 - z T}, \\
 \rho(T) = \frac{1}{\pi T^2} \mathop{\mathrm{Im}} \mathcal{F} (1/T + i 0). \label{rhoF}
\end{gather}
Note that the function $\mathcal{F}(z)$ has singularities when $z$ coincides with an inverse eigenvalue of $t^\dagger t$. That is why Eq.\ (\ref{rhoF})
involves an infinitesimal shift $i0$ in the argument of $\mathcal{F}(z)$.

It is often more convenient to use alternative variables $\theta$ and $\lambda$ defined as
\begin{equation}
\label{zlambda}
 z = \sin^2 \frac{\theta}{2},
\qquad
 T = \frac{1}{\cosh^2 \lambda},
\end{equation}
and to work in terms of the new generating function $\mathcal{F}(\theta)$ and distribution function $\rho(\lambda)$:
\begin{equation}
  \mathcal{F}(\theta) = \mathcal{F}(z) ,
\qquad
  \rho(\lambda) = \left|\frac{dT}{d\lambda}\right| \rho(T) .
\label{rholambda}
\end{equation}
The parameter $\lambda$ is referred to as the Lyapunov exponent. In terms of $\lambda$, the distribution function $\rho(\lambda)$ is constant for a diffusive
wire.\cite{Dorokhov84} The angle variable $\theta$ naturally appears in the sigma-model description of a disordered system as will be discussed in the following. In terms 
of $\theta$ and $\lambda$, relation (\ref{rhoF}) can be written as
\begin{equation}
 \label{TQ_rho}
 \rho(\lambda) = \frac{\sinh 2\lambda}{\pi} \mathop{\mathrm{Im}} \mathcal{F} (\theta = \pi + 2i\lambda - 0).
\end{equation}

\subsection{Matrix Green's function formalism}

Transport properties of a quasi-1D system can be studied within the matrix Green's function formalism developed by Nazarov in Ref.\ \onlinecite{Nazarov94}.

The moments of transmission distribution can be rewritten in terms of the retarded and advanced Green functions of the system with the help of the identity
\begin{equation}
\label{MGFF_trtn}
\tr (t^{\dagger} t)^n = \tr [\hat{v} G^A(x_L,x_R) \hat{v} G^R(x_R,x_L)]^n ,
\end{equation}
where $\hat{v}$ is the velocity operator and $x_L$ and $x_R$ are points in the left and right leads respectively. The relation (\ref{MGFF_trtn}) represents
the correspondence between Landauer\cite{Landauer57,Landauer70,Buttiker85} and Kubo description of electron transport. It allows to express the generating 
function $\mathcal{F}(z)$ in terms of a single matrix Green's function\cite{Nazarov94} defined as
\begin{equation}
\label{MGFF_GN}
 \begin{pmatrix}
   \epsilon - \hat{H} + i 0 & \sqrt{z} \hat{v} \delta(x - x_L) \\
   \sqrt{z} \hat{v} \delta(x - x_R) & \epsilon - \hat{H} - i 0
 \end{pmatrix} \check G(x,x')
  = \delta(x - x') \check{\mathbbm{1}}.
\end{equation}
Here, the Hamiltonian $\hat{H}$ acts in both real space and channel space and $\check{\mathbbm{1}}$ in the right-hand side is unity in the channel and matrix
retarded-advanced (RA) spaces. The off-diagonal terms in Eq.\ (\ref{MGFF_GN}) contain velocity operators and the parameter $z$. These two terms are located in
the right and left leads, which is ensured by the delta functions.

The generating function $\mathcal{F}(z)$ is related to the matrix Green's function $\check G$ in the following way:
\begin{equation}
\label{MGFF_Fz}
\mathcal{F}(z) = \frac{\partial}{\partial z} \ln Z, \qquad
Z(z) = \det \check G = \frac{\text{const}}{\det(1 - z t^{\dagger} t)}.
\end{equation}
The numerator of the last expression contains an irrelevant constant independent of $z$.

We will calculate transport properties of the system averaged over disorder realizations. This is most easily achieved with the help of the supersymmetric
representation. Consider two matrix Green functions (\ref{MGFF_GN}) with different source parameters $z_B$ and $z_F$ and define the partition function as
\begin{equation}
 \label{ZBF}
 Z(z_B, z_F)
  = \frac{\det \check G(z_B)}{\det \check G(z_F)} = \frac{\det(1 - z_F t^{\dagger} t)}{\det(1 - z_B t^{\dagger} t)}.
\end{equation}
This quantity can be viewed as a superdeterminant if the Green function $\check G$ is extended into a superspace. The two parameters $z_{B,F}$ have the meaning
of bosonic and fermionic source fields.

The generating function for transport characteristics can now be defined as an ordinary rather than logarithmic derivative
\begin{equation}
 \label{FzBzF}
 \mathcal{F}(z) = -\left. \frac{\partial Z(z_B, z_F)}{\partial z_F} \right|_{z_B = z_F = z}.
\end{equation}
This is made possible due to the supersymmetry condition $Z(z,z) = 1$. Owing to the linear relation between $\mathcal{F}$ and $Z$, we can directly average the
supersymmetric partition function over disorder. Such averaging is carried out within the non-linear sigma model as discussed in Sec.\ \ref{NLSM}.

Let us define the free energy corresponding to the disorder-averaged partition function as
\begin{equation}
 \label{Omega}
 \Omega(z_B, z_F) = - \ln \langle Z(z_B, z_F) \rangle.
\end{equation}
This free energy contains information on the full counting statistics of the electron transport in the system.\cite{Lee96, Levitov93} Similarly to Eq.\ 
(\ref{FzBzF}), it yields the average generating function
\begin{equation}
\label{avF}
\langle \mathcal{F}(z) \rangle = \left. \frac{\partial \Omega(z_B, z_F)}{\partial z_F} \right|_{z_B = z_F = z}.
\end{equation}
This function can be also written in terms of the bosonic and fermionic angular variables [cf.\ Eq.\ (\ref{zlambda})] defined as
\begin{gather}
 \label{zthetaBF}
 z_F = \sin^2 \frac{\theta_F}{2}, \qquad z_B = -\sinh^2 \frac{\theta_B}{2}, \\
 \label{avFtheta}
 \mathcal{F}(\theta) = \left. \frac{2}{\sin \theta} \frac{\partial \Omega(\theta_B,\theta_F)}{\partial \theta_F} \right|_{i \theta_B = \theta_F = \theta}.
\end{gather}
The average distribution function $\rho(\lambda)$, Eq.\ (\ref{TQ_rho}), can be expressed directly in terms of the free energy $\Omega$ as
\begin{equation}
 \label{rhotheta}
 \rho(\lambda)
  = \frac{2}{\pi} \mathop{\mathrm{Re}}
      \left. \frac{\partial \Omega(\theta_B, \theta_F)}{\partial \theta_F} \right|_{i \theta_B = \theta_F = \pi + 2i\lambda - 0}.
\end{equation}

The supersymmetric representation of the partition function $Z$, as well as the corresponding free energy $\Omega$, also allows us to access the mesoscopic 
fluctuations of conductance\cite{Lee85, Lee87, Mirlin94}
\begin{align}
\label{TQ_varG}
 \var G
  &= \langle G^2 \rangle - \langle G \rangle^2
  = G_0^2\, \left. \frac{\partial^2 \Omega(z_B, z_F)}{\partial z_F \partial z_B} \right|_{z_B = z_F = 0} \nonumber \\
  &= -4G_0^2\, \left. \frac{\partial^4 \Omega(\theta_B, \theta_F)}{\partial^2 \theta_B\; \partial^2 \theta_F} \right|_{\theta_B = \theta_F = 0}.
\end{align}
Variance of conductance describes correlations between different transmission channels. Hence, this quantity provides an additional information on the electron
transport not contained in the average generating function $\mathcal{F}$.

\subsection{Supersymmetric non-linear sigma model}
\label{NLSM}

In this section, we derive the disorder-averaged supersymmetric partition function (\ref{ZBF}) within the non-linear sigma model. We will provide detailed 
derivation for the problem of imbalanced quantum Hall edge (see Fig.\ \ref{QH}), that is described by the supersymmetric sigma model of the unitary symmetry 
class. Then, we discuss the case of the quantum spin Hall edge (symplectic class) and other symmetry classes.

\subsubsection{Quantum Hall edge (class A)}

Our derivation of the sigma model closely follows the steps of Refs.\ \onlinecite{Efetov99} and \onlinecite{Rejaei96} allowing for an unequal number of right- 
and
left-propagating modes $n_R \neq n_L$. We treat the edge states of the quantum Hall sample, Fig.\ \ref{QH}, assuming metallic limit with a short Fermi wave
length. This enables the quasiclassical description in terms of well defined conducting channels.\cite{Landauer57, Landauer70, Buttiker85} The Hamiltonian can 
be written as a matrix of
the size $N = n_R + n_L$,
\begin{equation}
\label{NLSM_H}
\hat{H} = - i \hat{v} \frac{\partial}{\partial x} + \hat V(x),
\end{equation}
where $\hat V(x)$ is the disorder potential and $\hat{v}$ is the velocity operator acting in the channel space,
\begin{equation}
\hat{v} = v \begin{pmatrix} \mathbbm{1}_{n_R} & 0 \\ 0 & -\mathbbm{1}_{n_L} \end{pmatrix}.
\end{equation}
We assume for simplicity that all channels have the same Fermi velocity $v$. The disorder potential $\hat V(x)$ is a random Hermitian matrix obeying Gaussian
distribution with $\langle \hat V \rangle = 0$ and
\begin{equation}
\langle V_{nm}(x) V_{mn}(x') \rangle = \frac{1}{N \tau}\, \delta(x - x') ,
\end{equation}
where $\tau$ is the electron mean free time.

The supersymmetric partition function $Z(z_B, z_F)$ can be written as a Gaussian integral over superfields,
\begin{gather}
\label{NLSM_ZFB}
Z(z_B, z_F) = \int D\psi^{\dagger}\, D\psi\; e^{-S[\psi^\dagger, \psi]}, \\
S = i \int dx\, \psi^{\dagger} \Lambda  \left( i \hat{v}\, \frac{\partial}{\partial x} - \hat V(x) + i 0 \Lambda + M \right) \psi. \label{Spsi}
\end{gather}
Here the supervector $\psi$ contains $2N$ complex and $2N$ Grassmann variables and operates in retarded-advanced (RA), Bose-Fermi (BF), and channel spaces. The
matrix
\begin{equation}
\label{Lambda}
\Lambda = \begin{pmatrix} 1 & 0 \\ 0 & -1 \end{pmatrix}_{\rm RA}
\end{equation}
operates in the RA space, while
\begin{equation}
M = \hat{v} \begin{pmatrix} \sqrt{z_B} & 0 \\ 0 & \sqrt{z_F} \end{pmatrix}_{\rm BF} \otimes \begin{pmatrix} 0 & \delta(x) \\ \delta(x-L) & 0 \end{pmatrix}_{\rm RA}
\end{equation}
represents the source terms in Eq.\ (\ref{MGFF_GN}). Gaussian integral in Eq.\ (\ref{NLSM_ZFB}) yields the superdeterminant of the corresponding matrix, that is
exactly the ratio of usual determinants from Eq.\ (\ref{ZBF}).

The term with the matrix $M$ implies that the eigenfunctions of the operator in the quadratic action (\ref{Spsi}) have jumps at the boundaries $x=0$ and $L$ 
between the sample and the leads. In order to eliminate these jumps and make the fields $\psi$ continuous across the boundaries, we perform a gauge rotation in 
the RA space,\cite{Rejaei96}
\begin{gather}
\psi = \Gamma \tilde \psi, \qquad \psi^{\dagger} = \tilde\psi^{\dagger} \Gamma^{-1}, \qquad \Lambda = \Gamma \tilde\Lambda \Gamma^{-1},
\label{gauge} \\
\hat v\, \frac{\partial \Gamma}{\partial x} = i M \Gamma. \label{EqGamma}
\end{gather}
The last equation ensures that $M$ drops from the action and determines $\Gamma$ up to right multiplication by a constant matrix.

We write the source fields $z_{B,F}$ in terms of the angles $\theta_{B,F}$ [see Eq.\ (\ref{zthetaBF})] and arrange them in a diagonal matrix
\begin{equation}
\hat{\theta} = \begin{pmatrix} i \theta_B & 0 \\ 0 & \theta_F \end{pmatrix}_{\rm BF}.
\end{equation}
In terms of the angle variables, Eq. (\ref{EqGamma}) can be explicitly solved yielding
\begin{equation}
\label{NLSM_G}
 \Gamma
  = \begin{dcases}
      \begin{pmatrix}
        i \cos (\hat{\theta}/2) & 0 \\
        0 & 1
      \end{pmatrix}_{\rm RA}\!\!, & x < 0, \\
      \begin{pmatrix}
        i \cos (\hat{\theta}/2) & i \sin (\hat{\theta}/2) \\
        0 & 1
      \end{pmatrix}_{\rm RA}\!\!, & 0 < x < L, \\
      \begin{pmatrix}
        i \cos (\hat{\theta}/2) & i \sin (\hat{\theta}/2) \\
        -\sin (\hat{\theta}/2) \cos (\hat{\theta}/2) & \cos^2 (\hat{\theta}/2)
      \end{pmatrix}_{\rm RA}\!\!, & x > L.
    \end{dcases}
\end{equation}

Once the off-diagonal source terms in the action (\ref{Spsi}) are removed, we proceed with the derivation of the sigma model in the standard
way.\cite{Efetov99} The next step is the averaging over disorder that yields the quartic term $(\tilde\psi^\dagger \tilde\psi)^2$. This term is decoupled by
a Hubbard-Stratonovich transformation introducing the supermatrix field $Q$ that acts in RA and BF spaces but not in the channel space. The action is then
integrated over $\tilde\psi$ and $\tilde\psi^\dagger$ leading to
\begin{multline}
\label{NLSM_trln}
S[Q] = \frac{N}{8 \tau} \Str Q^2
+ n_R \Str \ln \left(i v\, \frac{\partial}{\partial x} + \frac{i Q}{2\tau} \right) \\
+ n_L \Str \ln \left(-i v\, \frac{\partial}{\partial x} + \frac{i Q}{2\tau} \right).
\end{multline}
Here, ``Str'' is the full operator supertrace over all spaces (BF, RA, channels) including integration in the real space. We are using boson-dominated convention
$\Str A = \Tr A_{BB} - \Tr A_{FF}$ as in Ref.\ \onlinecite{Mirlin00}.

\begin{figure}
\center
\tabcolsep=3mm
\begin{tabular}{cc}
\includegraphics[height=0.18\textwidth]{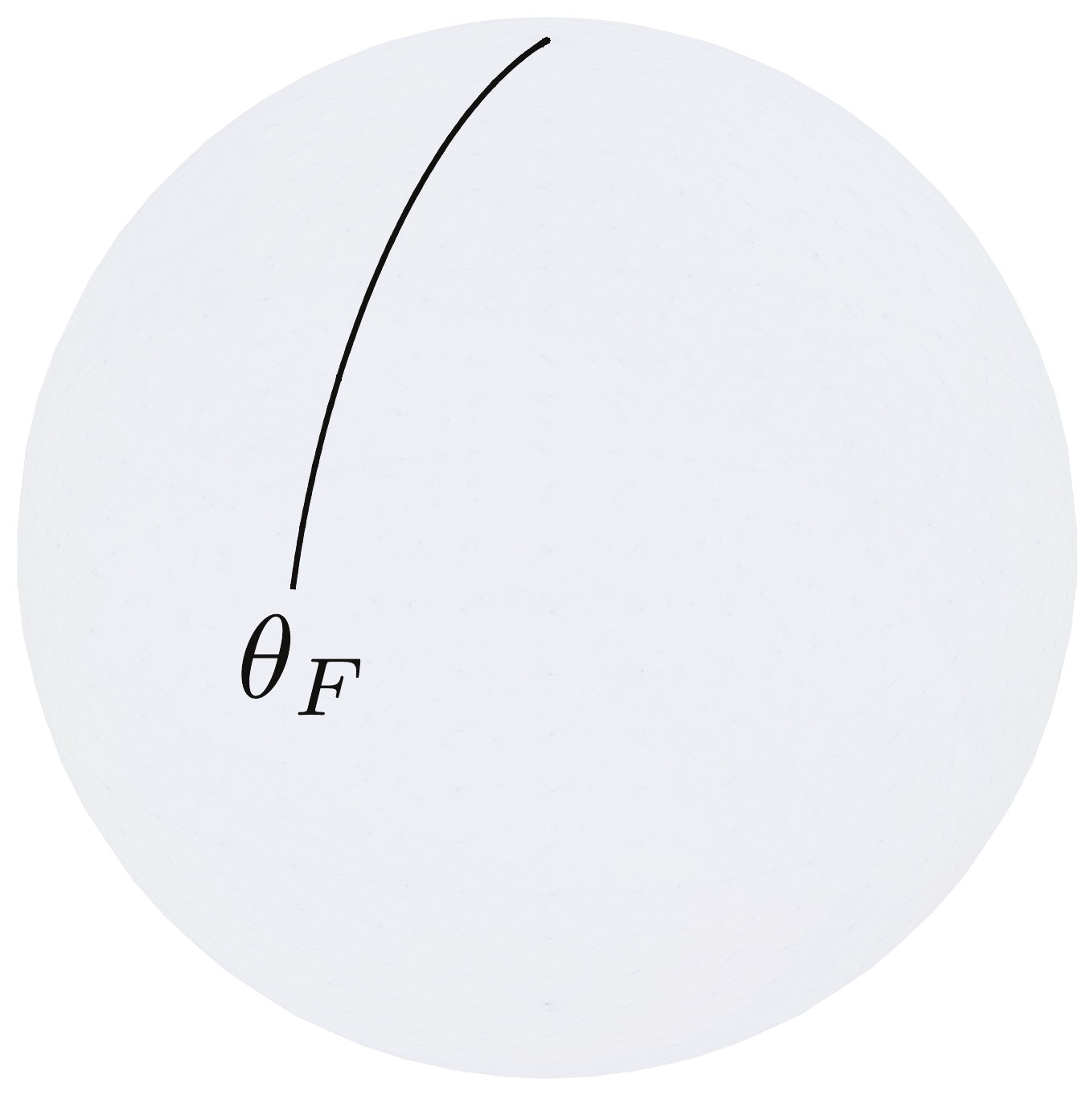} & \includegraphics[height=0.18\textwidth]{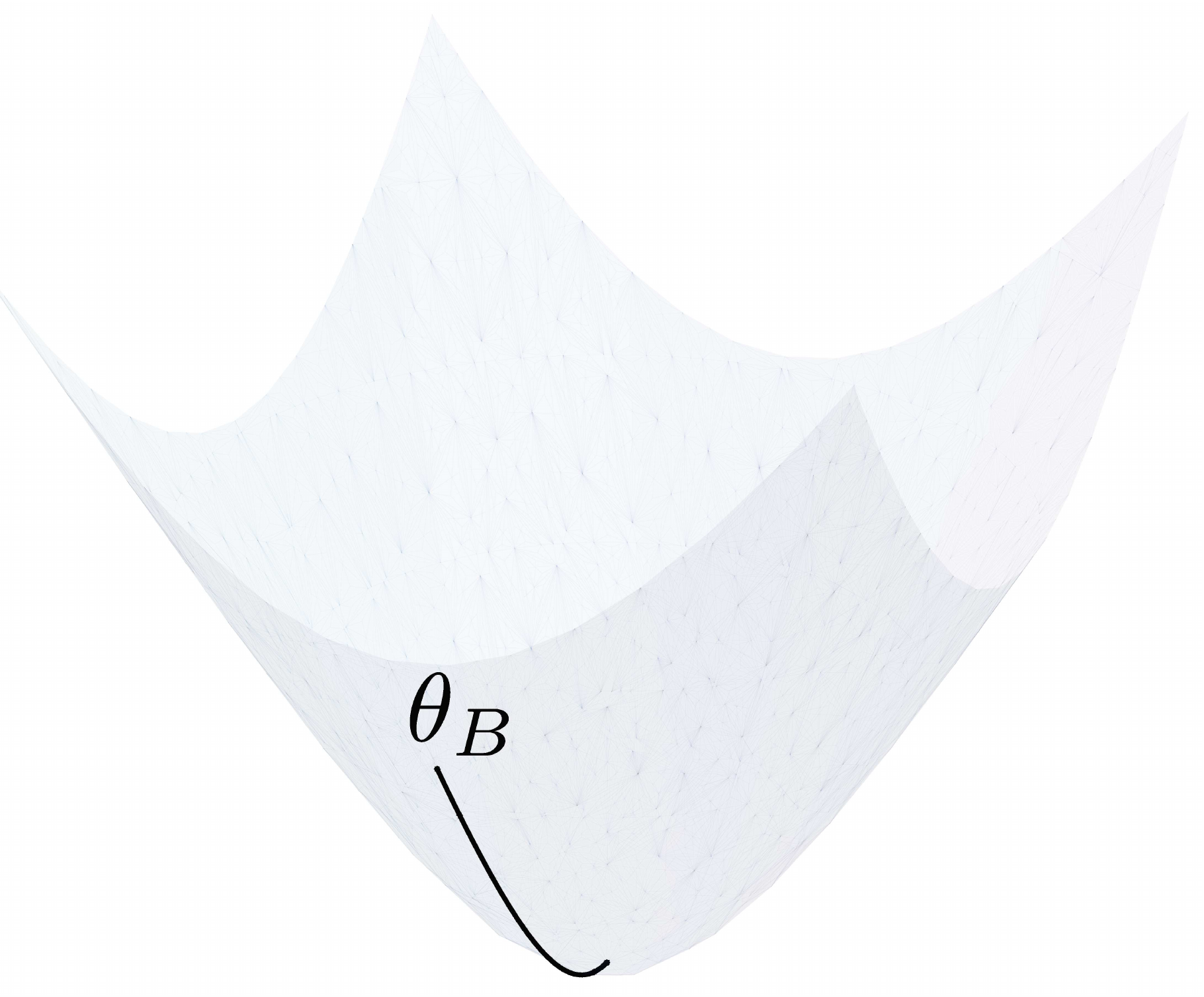}\\
FF & BB
\end{tabular}
\caption{Compact (FF) and non-compact (BB) sectors of the sigma-model manifold in the unitary symmetry class. The shortest geodesic trajectory connecting the
two boundary values (\ref{BC}) is shown.}
\label{UnitaryManifold}
\end{figure}

In the limit of large number of channels $N \gg 1$, we can treat the action (\ref{NLSM_trln}) within the saddle-point approximation.\cite{Efetov99} Assuming the
matrix $Q$ is constant in space, we identify a degenerate minimum of the action $Q = T^{-1} \Lambda T$ with any supermatrix $T$ acting in BF and RA spaces.
Convergence of the $Q$ integral at this saddle manifold is guaranteed by a suitable choice of the structure of $T$ in the complex plane.\cite{Mirlin00} This
implies a compact group manifold $\mathrm{U}(2)$ for the fermionic sector of the model (FF block of $T$) and non-compact $\mathrm{U}(1,1)$ group in the bosonic
sector (BB block of $T$). Together with Grassmann variables (BF and FB blocks), this represents the unitary supergroup $\mathrm{U}(1,1|2)$. The matrix $Q =
T^{-1} \Lambda T$ is invariant under rotations $T \mapsto K T$ if the matrix $K$ commutes with $\Lambda$. Thus, the actual configuration space of the sigma model
is the coset $\mathrm{U}(1,1|2) / \mathrm{U}(1|1) \times \mathrm{U}(1|1)$. Its compact (FF) and non-compact (BB) parts have the form of a sphere $S^2$ and a
hyperboloid $H^2$, respectively, as illustrated in Fig.\ \ref{UnitaryManifold}. This is the sigma-model manifold for a system of the unitary symmetry class A.

The effective low-energy theory is derived by a gradient expansion of Eq.\ (\ref{NLSM_trln}) assuming that $T(x)$ varies slowly in space. With a cyclic 
permutation of matrices under the supertrace, we recast the action in the form
\begin{multline}
\label{gradexp}
S[Q] = n_R \Str \ln \left(i v\, \frac{\partial}{\partial x} - i v \dot T T^{-1} + \frac{i \Lambda}{2\tau} \right) \\
+ n_L \Str \ln \left(-i v\, \frac{\partial}{\partial x} + i v \dot T T^{-1} + \frac{i \Lambda}{2\tau} \right).
\end{multline}
Here, we use the ``dot'' notation for derivative: $\dot T = \partial T/\partial x$.

Permutation of matrices, which was used in deriving Eq.\ (\ref{gradexp}), is equivalent to the rotation of the fields $\psi \mapsto T^{-1} \psi$,
$\psi^\dagger \mapsto \psi^\dagger T$ in the original action (\ref{Spsi}). Such a rotation may give rise to the chiral anomaly\cite{Fujikawa04, Bell69, 
Altland02} due to non-trivial Jacobian of the transformation. In the imbalanced case $n_L \neq n_R$, the anomalous contributions from the two terms of Eq.\ 
(\ref{gradexp}) do not cancel. Hence, in order to get rid of such contributions, we will assume $\sdet T = 1$.

Expanding the logarithms in Eq.\ (\ref{gradexp}) up to the second order in small derivatives $\dot T T^{-1}$, we obtain the action of the sigma model,
\begin{gather}
\label{NLSM_S}
 S[Q]
  = -\int_0^L dx\, \str \left[
      \frac{\xi}{8}\, {\dot Q}^2 + \frac{m}{2}\, T^{-1} \Lambda \dot T
    \right],\\
 \xi = N v \tau, \qquad m = n_R - n_L, \qquad Q = T^{-1} \Lambda T.
\end{gather}
Here, $\xi$ has the meaning of the localization length.

The second term in the action (\ref{NLSM_S}) is the 1D version of the topological Wess-Zumino-Witten term.\cite{Wess71,Witten83,Witten84} It appears due to the 
imbalance between right- and left-moving channels. Since the topological term is written explicitly in terms of the matrix $T$ rather than $Q$, we need to 
justify its gauge symmetry. Any transformation $T \mapsto KT$ with $[\Lambda, K] = 0$ and $\sdet K = 1$ leaves the matrix $Q$ invariant and maintains the 
condition $\sdet T = 1$ (cf.\ discussion of the anomaly above). The topological term changes under such transformation by the integral of the total $x$ 
derivative of $(m/2) \str\ln(K_R K_A^{-1})$ (here $K_{R,A}$ are the two diagonal blocks of the matrix $K$ in RA space), which equals $m \str\ln K_R$ due to the 
condition $\sdet K = 1$. The value of this integral is an integer multiple of $2\pi i$ for any closed trajectory $T(0) = T(L)$ provided $m$ is integer. Since 
the imbalance $m = n_R - n_L$ is an integer by definition, the weight $e^{-S}$ is a function of $Q$ only for any closed path $Q(x)$ on the sigma-model manifold.

Let us now establish the boundary conditions for the action (\ref{NLSM_S}). Since we have removed $M$ from Eq.\ (\ref{NLSM_ZFB}), all the fields, including the
matrix $Q$, are continuous at $x=0$ and $L$. Inside metallic leads, the matrix $Q$ takes the value $Q = \tilde\Lambda$ that results from the gauge rotation
(\ref{gauge}). Using (\ref{NLSM_G}), we obtain
\begin{equation}
\label{BC}
Q(0) = \Lambda, \qquad Q(L) = \begin{pmatrix} \cos\hat\theta & \sin\hat\theta \\ \sin\hat\theta & -\cos\hat\theta \end{pmatrix}_{\rm RA}.
\end{equation}
Thus we see that the source parameters $\theta_{B,F}$ enter the sigma model only via the boundary conditions. The partition function (\ref{ZBF}) is given by
the path integral in the superspace of the matrix $Q$ with the action (\ref{NLSM_S}). All the paths connect the point $Q = \Lambda$, representing the
``north pole'' of the sphere in the compact FF sector and the base of the hyperboloid in the non-compact BB sector, with a point described by the polar angles
$\theta_F$ and $\theta_B$ on the sphere and the hyperboloid, respectively, as shown in Fig.\ \ref{UnitaryManifold}.

In view of the boundary conditions (\ref{BC}), we have to reconsider the gauge invariance of the topological term. As was argued above, the gauge symmetry is
preserved only for closed paths, which contradicts the boundary conditions (\ref{BC}). In order to obtain gauge-invariant expressions, we should ``close the
circuit'' by adding the contribution of a specific trajectory going from the final point back to the initial point. Choosing this backward path to be the
shortest (geodesic) restores gauge invariance and guarantees that the conductance $G$ obtained from the model has the expected form (\ref{Gtotal}).

\subsubsection{Quantum spin Hall edge (class AII)}

Derivation of the sigma model for the edge of the quantum spin Hall system (Fig.\ \ref{QSH}) is similar to the case of the quantum Hall edge discussed above.
The main difference arises due to an additional spin degree of freedom and the associated time-reversal symmetry of the symplectic type:
\begin{equation}
 H = s_y H^T s_y.
 \label{Tsym}
\end{equation}
Here, $s_y$ is the Pauli matrix acting on the electron spin. We can describe the edge modes by the 1D Hamiltonian (\ref{NLSM_H}) assuming that the channels are
fully spin-polarized (right- and left-moving states have spin up and down respectively). The number of right- and left-propagating channels is the same,
$N = n_R = n_L$, and the velocity operator $\hat{v} = v \mathbbm{1}_N \otimes s_z$.

The partition function $Z(z_B, z_F)$ is represented by an integral over the supervector $\psi$ as in Eq.\ (\ref{NLSM_ZFB}) and the source fields are removed by
the gauge transformation (\ref{gauge}). The time-reversal symmetry is then taken into account explicitly by doubling the size of the matrix introducing an
extra time-reversal (TR) space,
\begin{gather}
\label{NLSM_SAII}
S = i \bar{\Psi} \left( i v s_z \frac{\partial}{\partial x} - V  + i 0 \Lambda \right) \Psi, \\
\Psi = \frac{1}{\sqrt{2}} \begin{pmatrix} \psi \\ i s_y \psi^* \end{pmatrix}_{\rm TR}\!\!, \quad
\bar{\Psi} = \frac{1}{\sqrt{2}} \Bigl( \psi^{\dagger} \Lambda,\; -i \psi^T s_y k \Lambda \Bigr)_{\rm TR} ,
\end{gather}
where $k = \diag\{1, -1\}_{\rm BF}$.
The derivation proceeds along the standard route\cite{Efetov99, Mirlin00} with disorder averaging, Hubbard-Stratonovich
transformation, and integrating over $\Psi$. The resulting action is similar to Eq.\ (\ref{NLSM_trln}):
\begin{equation}
\label{AII_trln}
S[Q] = \frac{N}{8 \tau} \Str Q^2
+ N \Str \ln \left(i v\, \frac{\partial}{\partial x} + \frac{i Q}{2\tau} \right).
\end{equation}
The matrix $Q$ operates in BF, RA, and TR spaces and has twice larger size as compared to the quantum Hall problem studied above. Apart from this, an
additional constraint on $Q$ appears due to the time-reversal symmetry:
\begin{equation}
\label{C}
Q = \bar Q = C^T Q^T C, \qquad C = \begin{pmatrix} i \sigma_y & 0 \\ 0 & \sigma_x \end{pmatrix}_{\rm BF}.
\end{equation}

\begin{figure}
\center
\includegraphics[width=0.4\textwidth]{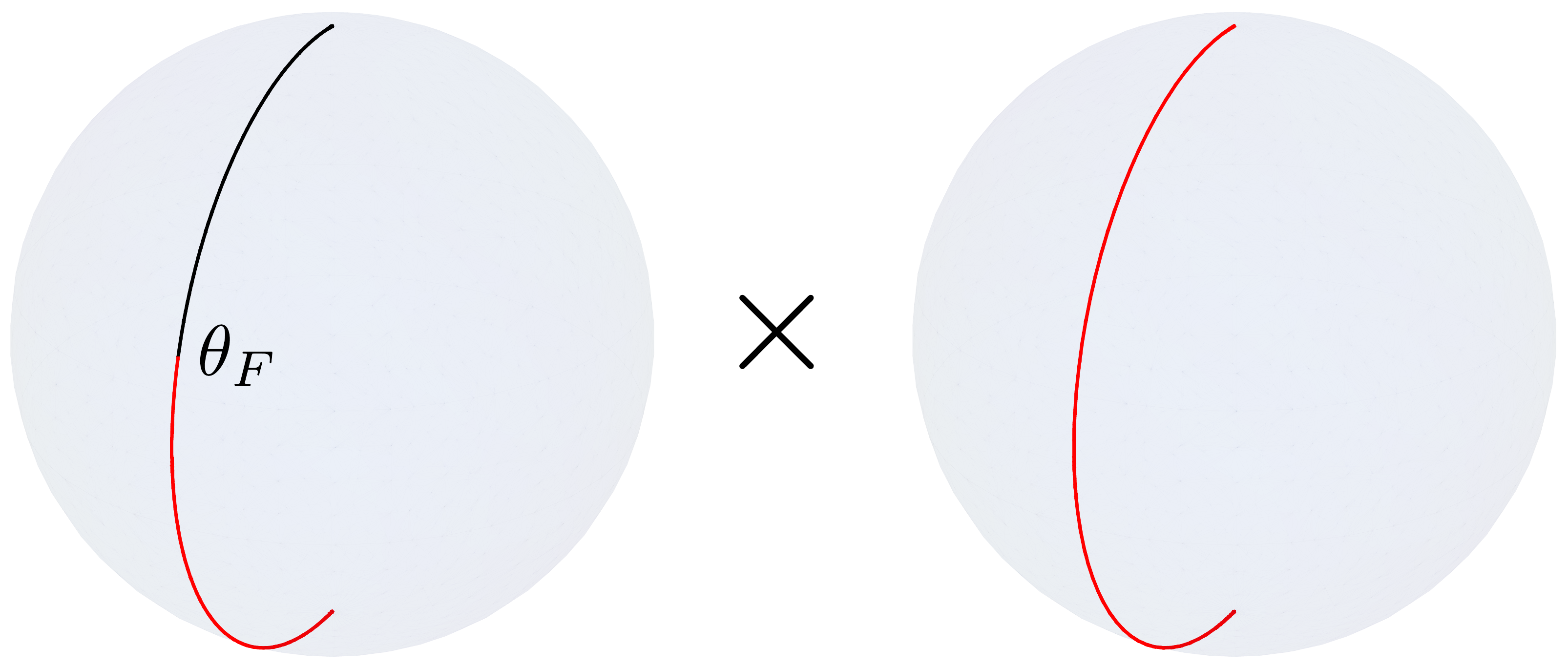}
\caption{Compact sector of the sigma-model manifold in the symplectic class. The space has the structure of a direct product of two spheres $S^2 \times
S^2/\mathbb{Z}_2$. Factorization by $\mathbb{Z}_2$ identifies the opposite points. Two topologically distinct trajectories between two fixed points are shown.}
\label{SymplecticManifold}
\end{figure}

In the limit $N \gg 1$, the saddle-point analysis of the action (\ref{AII_trln}) yields $Q = T^{-1} \Lambda T$. The condition (\ref{C}) is fulfilled if
\begin{equation}
\label{K}
T^{-1} = \bar T K, \qquad [\Lambda, K] = 0.
\end{equation}
The matrix $K$ can be arbitrary since $Q$ does not depend on it. The standard choice is $K = 1$ and $T^{-1} = \bar T$. In this case, the matrix $T$ belongs to
the orthogonal group $\mathrm{O}(4)$ in the compact sector (FF block) and to the group $\mathrm{Sp}(2,2)$ in the non-compact sector (BB block). The
matrix $Q$ remains invariant under left rotations of $T$ with any matrix that commutes with $\Lambda$. Thus, the compact sector of the model is given by the
coset space $\mathrm{O}(4)/\mathrm{O}(2) \times \mathrm{O}(2)$ and the non-compact sector is $\mathrm{Sp}(2,2)/\mathrm{Sp}(2) \times \mathrm{Sp}(2)$. The former
is the four-dimensional manifold with the structure of a direct product of two spheres $S^2 \times S^2/\mathbbm{Z}_2$ as shown in Fig.\
\ref{SymplecticManifold}. Factorization by $\mathbbm{Z}_2$ implies that simultaneous inversion of both spheres yields the same value of $Q$. The non-compact
sector of $Q$ has the structure of a four-dimensional hyperboloid $H^4$. The sigma-model manifold for a system of the symplectic symmetry class AII includes
these compact and non-compact sectors along with Grassmann variables connecting them.

Gradient expansion in Eq.\ (\ref{AII_trln}) is carried out similarly to the case of the quantum Hall problem and yields the action
\begin{equation}
 S[Q]
  = -\int_0^L dx\, \str \left[
      \frac{\xi}{8}\, {\dot Q}^2 + \frac{N}{2}\, T^{-1} \Lambda \dot T
    \right],
\end{equation}
with the same boundary conditions (\ref{BC}). However, the topological term in this action has different properties. As was argued above, gauge symmetry of
the topological term is ensured only for closed trajectories in terms of the matrix $T$. The continuity of $T$ is not always compatible with the standard
convention $\bar T T = 1$. Therefore, considering the topological term, we will allow for a non-trivial matrix $K$ in Eq.\ (\ref{K}), which also continuously
changes along the trajectory with $\sdet K = 1$. The topological term can be transformed in the following way:
\begin{align}
\label{NLSM_SWZW}
 S_\text{top}
  &= -\frac{N}{2} \int dx \str T^{-1} \Lambda \dot T
  = -\frac{N}{2} \int dx \str \dot {\bar T} \Lambda \bar T^{-1} \nonumber \\
  &= \frac{N}{2} \int dx \str [T^{-1} \Lambda \dot T + K^{-1} \Lambda \dot K] \nonumber \\
  &= -S_\text{top} + N \int dx \frac{\partial}{\partial x} \ln \sdet K_R,
\end{align}
where $K_R$ is the upper diagonal block of $K$ in RA space. We thus see that possible values of the topological term are integer multiples of $i \pi N$.

For a sample with an even number of edge channels, the topological term is ineffective and can be dropped from the action. However, when $N$ is odd, some
trajectories will contribute to the partition function with a negative sign. Since the topological term takes only discrete values, its variation vanishes.
Hence, only topologically distinct trajectories can yield different values of $S_\text{top}$.

The compact sector of the sigma-model manifold is doubly connected with the homotopy group $\pi_1 = \mathbbm{Z}_2$. This is illustrated in Fig.\ 
\ref{SymplecticManifold}. A trajectory going from the two ``north poles'' of both spheres to the two ``south poles'' is a closed trajectory in terms of $Q$. 
This trajectory cannot be continuously shrunk to a single point, hence, it is a representative of the non-trivial homotopy class. An explicit calculation of the 
action along such a trajectory yields $S_\text{top} = i \pi N$. This proves that the action does contain the $\mathbbm{Z}_2$ topological term whenever the 
number of channels is odd.

\subsubsection{Other symmetry classes}

Non-trivial topological terms may also arise in the sig\-ma-model action of a system with other symmetries.\cite{Kitaev09, Schnyder09} The $\mathbb{Z}$-type
topology occurs in classes C and D that are superconducting analogs of the unitary class A. Class C refers to a superconductor with broken time-reversal but
preserved spin symmetry, while class D implies both symmetries broken.\cite{Altland97} These classes describe the edge of spin and thermal quantum Hall
sample, respectively. Corresponding sigma-model action has the form of Eq.\ (\ref{NLSM_S}) with the boundary condition (\ref{BC}) and an additional constraint
\begin{equation}
\bar{Q} = C^T Q^T C = - Q,
\end{equation}
that occurs due to the specific particle-hole symmetry of the superconductor. The matrix $C$ is skew symmetric in the FF sector and symmetric in the BB sector
in class C and vice versa for class D.

The analog of the quantum spin Hall system is provided by the symmetry class DIII. This is a superconductor with preserved time reversal but broken spin
symmetry.\cite{Altland97, Lamacraft04} The system possesses $\mathbbm{Z}_2$ topological properties and represents a possible topological superconductor in
2D.\cite{Kitaev09, Schnyder09} The corresponding sigma model is defined on the group manifold with the compact sector being the group $O(2)$. The
derivation is quite similar to the case of the symplectic class and the topological term emerges whenever the number of edge channels is odd.

\section{Electron transport at the edge of $\mathbbm{Z}$ topological insulators}
\label{QA}

In this section, we consider the general transport characteristics of the edge of a $\mathbbm{Z}$ topological insulator. Main calculations will be carried out 
for the imbalanced quantum Hall edge (unitary class A), as depicted in Fig.\ \ref{QH}. The results for spin and thermal quantum Hall system (classes C and D, 
respectively) will be discussed in the end of the section. We will assume the limit of a short sample $L \ll \xi$, when the localization effects are weak.
Nevertheless, taking account of the topological term strongly modifies transport properties even in this semiclassical case.

\subsection{Distribution of transmission probabilities}

The distribution of transmission probabilities at an imbalanced quantum Hall edge can be derived from the sigma model (\ref{NLSM_S}) with twisted boundary
conditions (\ref{BC}), as explained in the previous section. For a short sample,
$L \ll \xi$, the path integral is dominated by the extrema of
the action and we can use the quasiclassical approximation
to solve the problem. Classical trajectories do not involve
Grassmann variables, hence, the compact (FF) and
noncompact (BB) sectors of the model decouple. This
means that the quasiclassical results can be derived from
a simplified sigma model involving only a fermionic (compact)
or a bosonic (noncompact) sector without Grassmann
variables. \cite{Campagnano06} In this section, we choose to work
with the compact version. 

\begin{figure}
\center
\includegraphics[width=0.2 \textwidth]{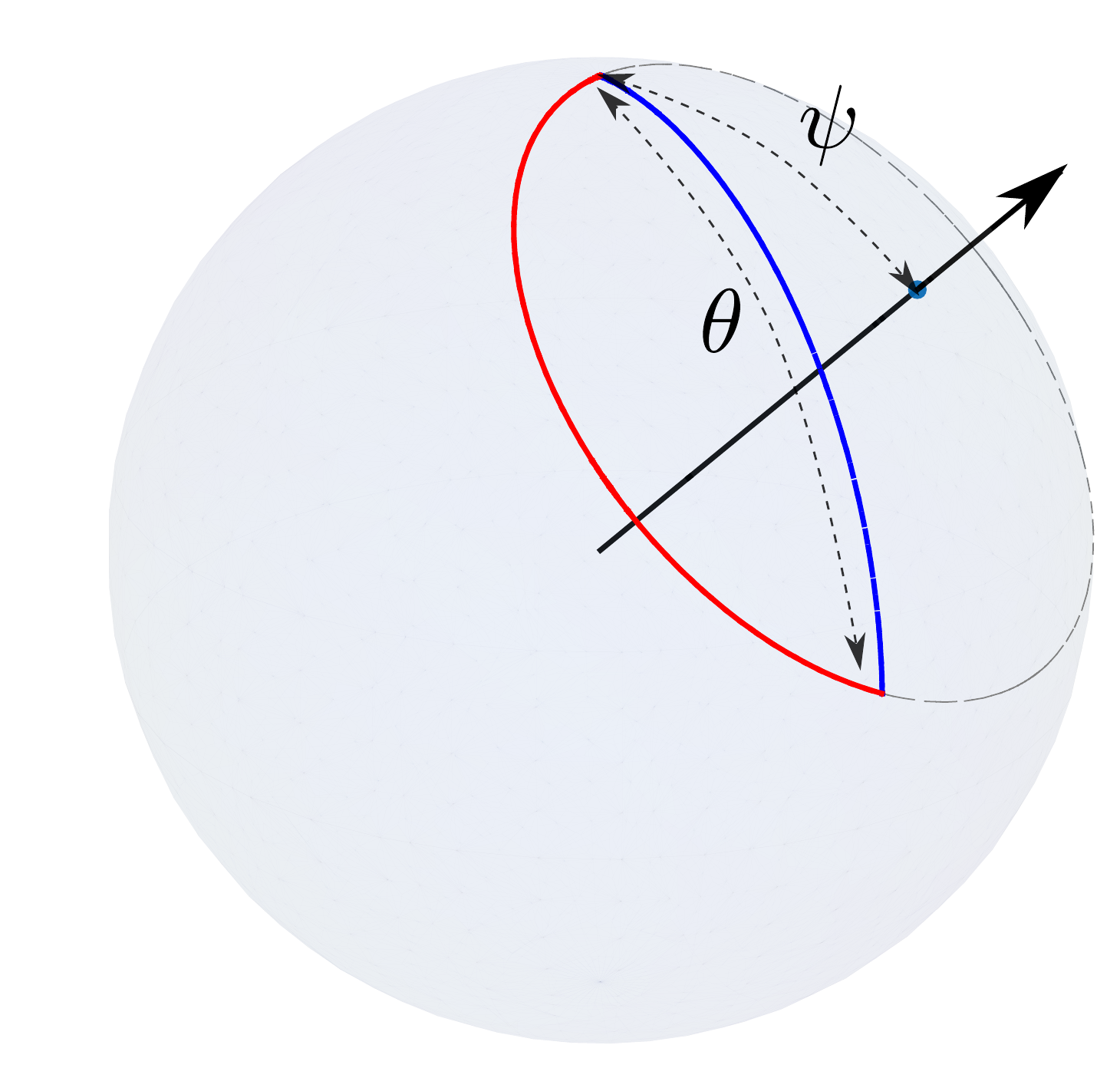}
\caption{Classical solution in the compact sector of the unitary sigma model in the presence of the topological term (\ref{smallerarc}). The trajectory starts
at the ``north pole'' and ends at the point with the polar angle $\theta$. The solution can be thought of as a rotation with a constant angular velocity
around the tilted axis $\psi$.}
\label{Trajectory}
\end{figure}

Let us consider the classical action for the compact sector of the sigma model. Parametrizing the sphere by the polar angle $\theta$ and the azimuthal angle
$\phi$, we can write the action as
\begin{equation}
\label{QA_S}
 S
  = \frac{\xi}{2L} \int_0^1 dx \left[
      \frac{1}{2}(\dot{\theta}^2 + \sin^2 \theta \, \dot{\phi}^2) - i \alpha (1 - \cos\theta) \dot{\phi}
    \right],
\end{equation}
where we have rescaled $x$ such that the integration interval extends up to $x = 1$, and the dimensionless parameter $\alpha$ is defined as
\begin{equation}
\alpha = \frac{m L}{\xi} = \frac{(n_R - n_L) L}{(n_R + n_L) l}.
\end{equation}
The action (\ref{QA_S}) describes the motion of a particle on a sphere in the magnetic field created by a monopole located in the center of the sphere. The 
topological term can be also interpreted as the Berry phase proportional to the solid angle encircled by the trajectory. In the absence of magnetic field, 
classical trajectories are arcs of great circles, i.e., geodesics on the sphere. Magnetic field exerts a Lorentz force on the moving particle, thus making its 
trajectory an arc of a smaller circle (Fig.\ \ref{Trajectory}).

A classical solution minimizing the action (\ref{QA_S}) satisfies the Euler-Lagrange equations
\begin{gather}
\frac{d}{d x} \left[ \sin^2 \theta \dot{\phi} + i \alpha \cos \theta \right] = 0, \\
\ddot{\theta} - \sin \theta \cos \theta \dot{\phi}^2 + i \alpha \sin \theta \dot{\phi} = 0.
\end{gather}
It is convenient to represent the classical trajectory in the rotated frame, as shown in Fig.\ \ref{Trajectory}. For the path starting at the north pole,
we select the polar axis tilted by the angle $\psi$. In such coordinates the trajectory is
\begin{equation}
\label{smallerarc}
 \theta' = \psi = \text{const}, \qquad \phi' = \chi x, \qquad \chi = \frac{i\alpha}{\cos\psi}.
\end{equation}
It represents a particle moving only in the azimuthal direction with a constant speed $\chi$ along a smaller circle defined by the constant value $\psi$. The
polar angle $\theta(x)$ in the original frame can be determined as the length along the arc of a great circle connecting the initial and current points of the
trajectory. Applying the law of cosines on the sphere, we obtain
\begin{equation}
\label{theta(x)}
 \sin\frac{\theta(x)}{2} = \sin \psi\; \sin \frac{\chi x}{2}.
\end{equation}
This solution should satisfy the final condition $\theta(1) = \theta_F$. Eliminating the constant $\psi$ with the help of Eq.\ (\ref{smallerarc}), we obtain a
transcendental equation for the angular velocity $\chi(\theta_F,\alpha)$:
\begin{equation}
 \label{QA_chitheta}
 \chi^2 \sin^2 \frac{\theta_F}{2} = (\chi^2 + \alpha^2) \sin^2 \frac{\chi}{2}.
\end{equation}
In the absence of the topological term, $\alpha = 0$, this equation yields $\chi = \theta_F$. In the imbalanced case, $\alpha \neq 0$, we assume that $\chi$ is 
an analytic function of $\theta_F$ in the region $0 < \re\theta_F < \pi$, $\im \theta_F > 0$ taking the value $\chi = i\alpha$ at $\theta_F = 0$. This 
assumption will allow us to relate classical dynamics on the sphere (for real $\theta_F$) and on the hyperboloid (positive imaginary $i\theta_B$) and to 
analytically continue both solutions to the positive values of $\lambda$, cf.\ Eq.\ (\ref{rhotheta}).

In the framework of quasiclassical approximation, the partition function is $Z = \exp[S_\text{min}(i\theta_B) - S_\text{min}(\theta_F)]$ with $S_\text{min}$ 
being the minimized classical action on the sphere. For the trajectory (\ref{smallerarc}), this action takes the value (up to a constant)
\begin{equation}
 \label{QA_Smin}
 S_\text{min}(\theta_F)
  = \frac{\xi}{L} \left[
      \frac{\chi^2}{4} + \alpha \arctanh\left( \frac{\alpha}{\chi} \tan\frac{\chi}{2} \right)
    \right],
\end{equation}
while $\chi$ is determined by Eq.\ (\ref{QA_chitheta}).

\begin{figure}
\center
\begin{tabular}{c}
\includegraphics[width=0.95\columnwidth]{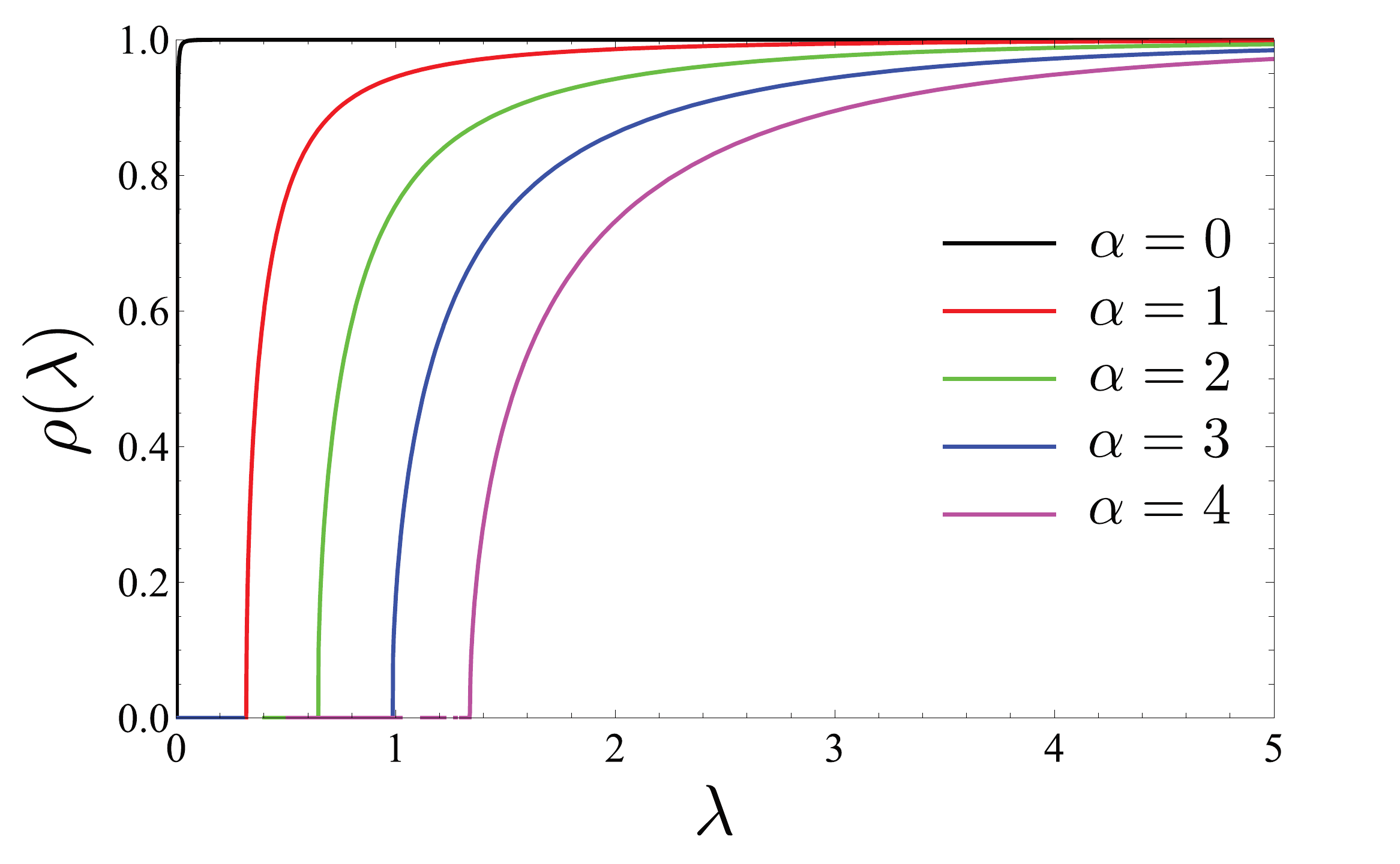} \\ \includegraphics[width=0.93\columnwidth]{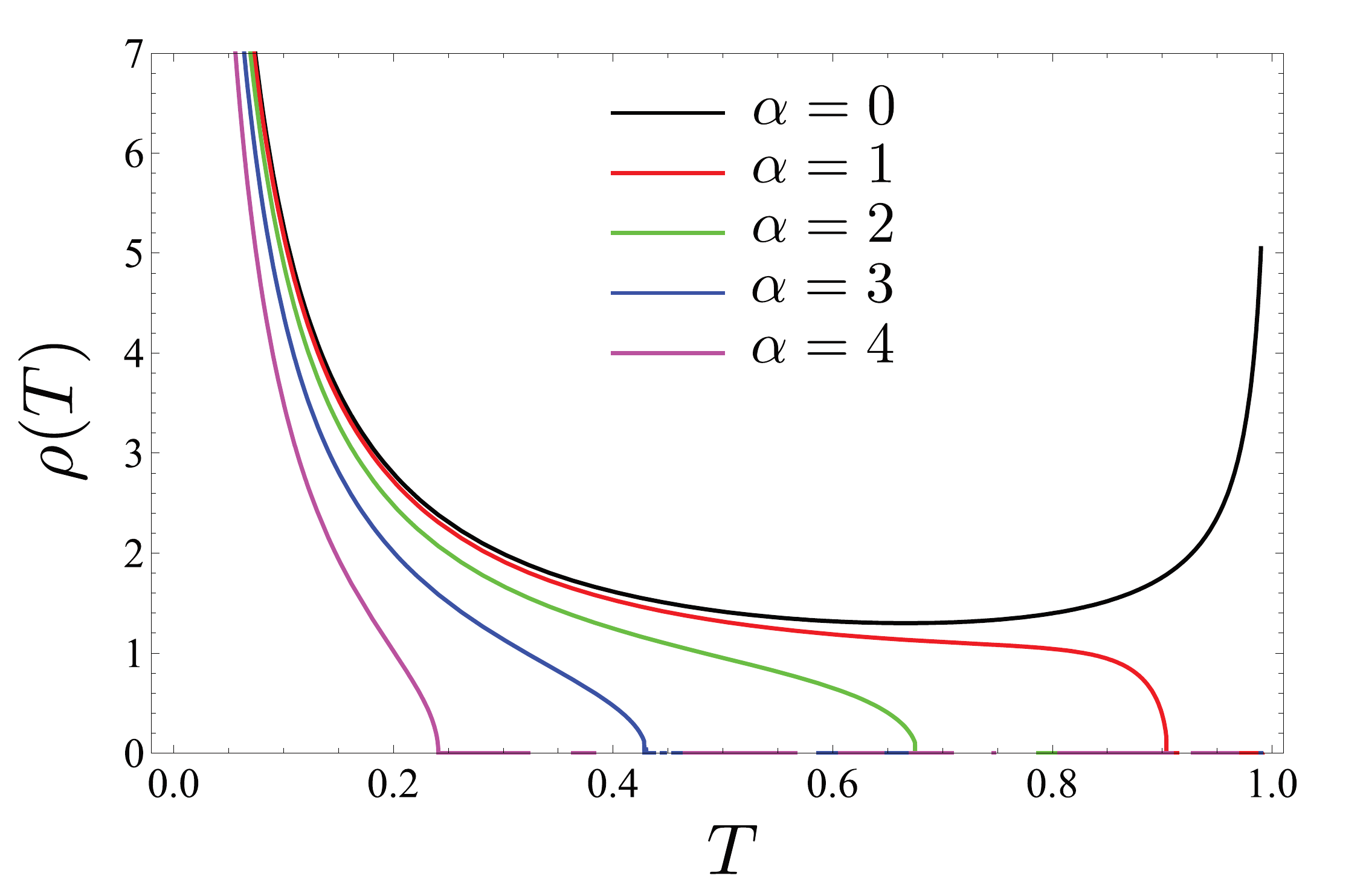}
\end{tabular}
\caption{Average distribution of transmission probabilities in the unitary class, Eq.\ (\ref{QA_rhol}) in terms of $\lambda$ (upper panel) and $T$ (lower
panel) for different values of the parameter $\alpha$. Topologically protected states yield a delta peak at $\lambda = 0$ ($T = 1$), Eq.\ (\ref{delta}), and
a semiclassical gap of the size $\lambda_g$ or $T_g$, see Fig.\ \ref{GapBeta}.}
\label{SemiClassicalRho}
\end{figure}

The generating function $\mathcal{F}(\theta)$ can be calculated from Eq.\ (\ref{QA_Smin}) with the help of Eq.\ (\ref{avFtheta}). It is, however, technically 
easier to use the relation between the derivative of the classical action and momentum, $\partial S_\text{min}/\partial \theta_F = (\xi/2L) \dot \theta(1)$, 
and apply Eq.\ (\ref{theta(x)}). This way we obtain the generating function
\begin{equation}
 \mathcal{F}(\theta)
  = \frac{\xi}{L \sin \theta}  \sqrt{\chi^2 + \frac{\alpha^2}{\cos^2(\theta/2)}}.
 \label{Ftheta}
\end{equation}
The average distribution function is then given by Eq.\ (\ref{TQ_rho}),
\begin{equation}
\label{QA_rhol}
\rho(\lambda) = \frac{\xi}{\pi L} \re \sqrt{\chi^2 - \frac{\alpha^2}{\sinh^2 (\lambda + i0)}}.
\end{equation}
In the limit $\alpha = 0$, this yields a constant $\rho(\lambda) = \xi/L$ corresponding to the celebrated Dorokhov distribution\cite{Dorokhov84} for disordered
wires. For a finite imbalance $\alpha > 0$, the resulting distribution is shown in Fig.\ \ref{SemiClassicalRho}.

The main qualitative feature of the imbalance between right- and left-moving modes is the appearance of a gap in the transmission probability distribution.
This gap develops around $T = 1$ or, equivalently, $\lambda = 0$. Emergence of the gap can be explained within the mechanical picture considered above.
Transmissions close to $T = 1$ correspond to the value of the source angle close to $\theta = \pi$. The probability to find a channel with large transmission
is related to the classical action for a particle going from the ``north pole'' of the sphere almost to its ``south pole'' in a given time. Since the
left-right imbalance implies a magnetic monopole in the center of the sphere, all classical trajectories are deviated by the Lorentz force and the south pole
becomes classically unreachable.

Another qualitative feature of the distribution (\ref{QA_rhol}) is the delta peak at $\lambda = 0$. It appears due to the infinitesimal $i0$ term that becomes
effective at small $\lambda$ and yields
\begin{equation}
\label{delta}
\rho(\lambda \rightarrow 0) = -\frac{\xi |\alpha|}{\pi L} \im \frac{1}{\lambda + i 0} = |m|\, \delta(\lambda).
\end{equation}
This delta peak describes $|n_R - n_L|$ perfectly conducting topologically protected channels at the imbalanced edge of the sample. The gap in the distribution
function is a result of the statistical repulsion of transmission probabilities from the delta peak at $T = 1$.

The critical value $\lambda = \lambda_g$, that determines the size of the gap, can be calculated as follows. Close to $\lambda_g$, the transcendental equation
(\ref{QA_chitheta}) has two almost degenerate solutions. This implies that at the critical value of $\theta_F$ the derivative $\partial\theta_F / \partial\chi$
vanishes, which yields the equation for the critical value $\chi_g(\alpha)$:
\begin{equation}
\label{QA_gap}
\tan \frac{\chi_g}{2} = \frac{\chi_g}{2} + \frac{\chi_g^3}{2\alpha^2}.
\end{equation}
For $\alpha < 2\sqrt{3}$, solution of this equation is real and lies in the interval $0 < \chi_g < \pi$. At larger $\alpha$, the critical value $\chi_g$ is 
imaginary.

\begin{figure}
\center
\begin{tabular}{cc}
\includegraphics[width=0.95\columnwidth]{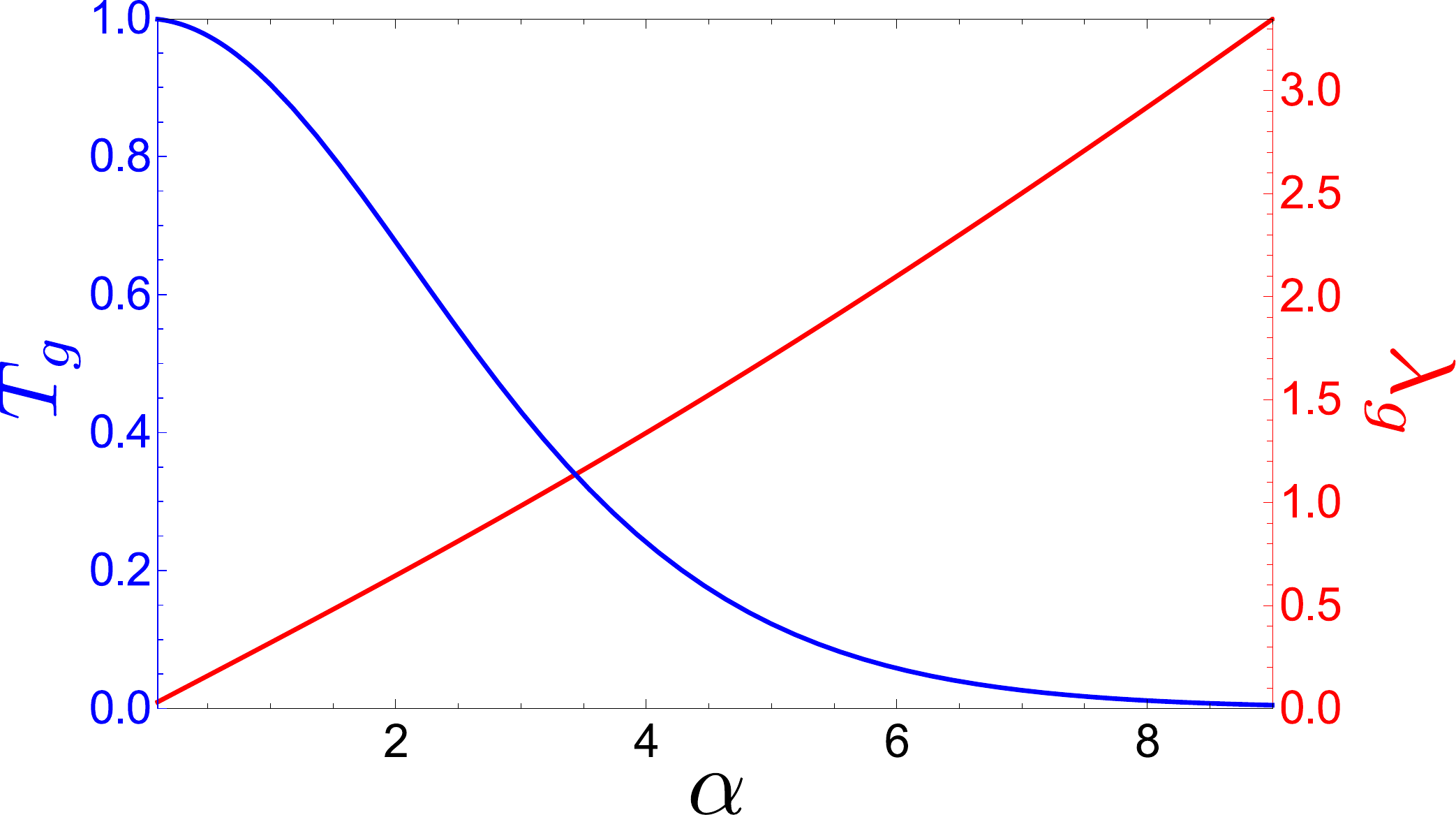}
\end{tabular}
\caption{The value of the gap in the average distribution function in terms of $T$ and $\lambda$ obtained by solving Eq.\ (\ref{QA_gap}). The asymptotics of
$\lambda_g$ are given in Eq.\ (\ref{QA_lambdag}).}
\label{GapBeta}
\end{figure}

Dependence of the gap on the parameter $\alpha$ is shown in Fig. \ref{GapBeta}. Its asymptotic behavior for small and large $\alpha$ can be derived by
approximately solving Eq.\ (\ref{QA_gap}) and using Eq.\ (\ref{QA_chitheta}).
\begin{equation}
\label{QA_lambdag}
\lambda_g = \begin{dcases}
\frac{\alpha}{\pi} + \left( \frac{2}{\pi^5} - \frac{1}{6 \pi^3} \right) \alpha^3 + \ldots, & \alpha \ll 1, \\
\frac{\alpha}{2} - \frac{\ln \alpha}{2} + \ldots, & \alpha \gg 1.
\end{dcases}
\end{equation}

Close to the critical value $\lambda_g$, the distribution function exhibits the square root behavior
\begin{equation}
\label{QA_rholg}
\rho(\lambda) = \frac{\xi}{\pi L}\, f(\alpha) \sqrt{\lambda - \lambda_g}, \qquad \lambda > \lambda_g,
\end{equation}
The value of the factor $f(\alpha)$ can be calculated by expanding Eq.\ (\ref{QA_chitheta}) close to the critical point determined by Eq.\ (\ref{QA_gap}). This
leads to
\begin{equation}
\label{QA_falpha}
 f(\alpha)
  = \left|\frac{2 \sqrt{\chi_g} (\chi_g - \sin\chi_g)^{3/4}}{\sqrt{2 \chi_g + \chi_g \cos \chi_g - 3\sin\chi_g} \sin^{1/4}\chi_g}\right|.
\end{equation}
This function behaves as $\sqrt{2\pi^3/\alpha}$ in the limit $\alpha \ll 1$ and approaches the value $2$ at large $\alpha$. It will be used in the discussion
of the universal crossover dependence of $\rho(\lambda)$ in the vicinity of $\lambda_g$ in Sec.~\ref{M0D}.

\subsection{Transport moments}
\label{QA_CFF}

\begin{figure}
\center
\begin{tabular}{c}
\includegraphics[width=0.95\columnwidth]{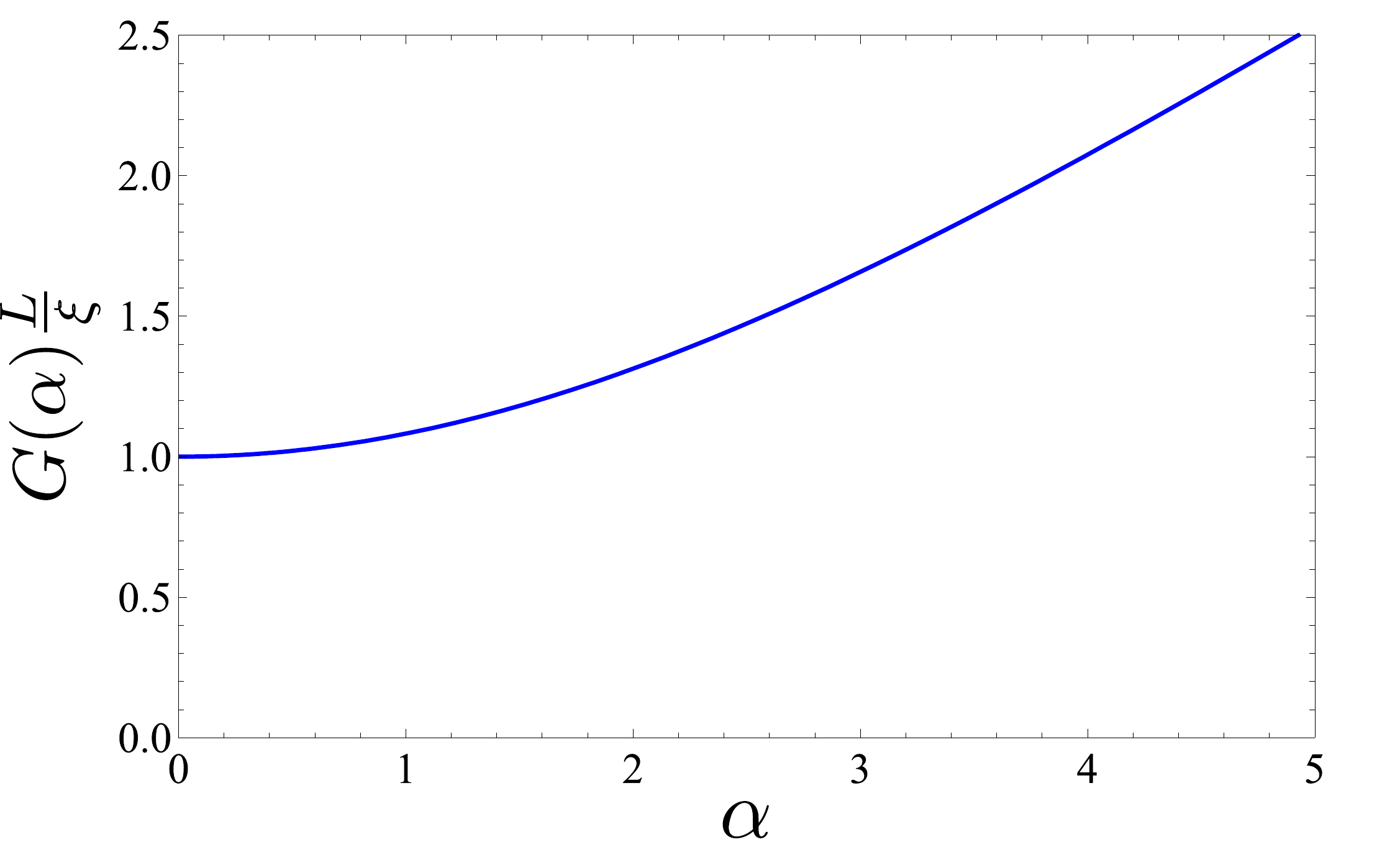} \\ \includegraphics[width=0.95\columnwidth]{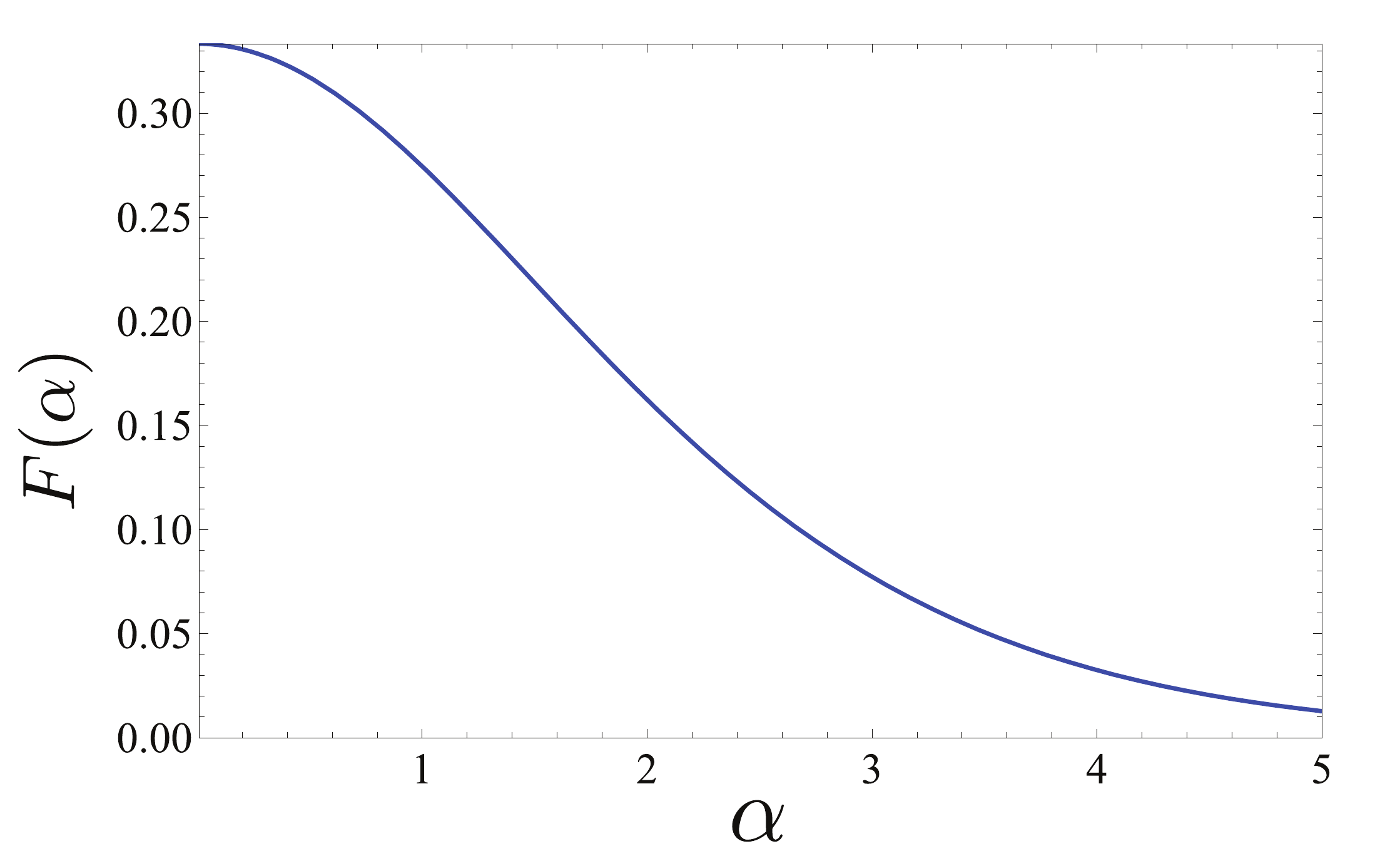}
\end{tabular}
\caption{Average conductance (upper panel) and Fano factor (lower panel) as a function of $\alpha$, Eqs.\ (\ref{QA_G}) and (\ref{QA_F}).}
\label{ConductanceFanoFactor}
\end{figure}

The distribution function of transmission probabilities allows to compute average conductance, Fano factor, and higher moments of electron transport. However,
the result (\ref{QA_rhol}) is written in terms of the parameter $\chi$ that is determined by the complicated transcendental equation (\ref{QA_chitheta}). It is
easier to compute the moments directly from the generating function (\ref{Ftheta}) by taking its derivatives at $\theta = 0$. We solve Eq.\ (\ref{QA_chitheta})
perturbatively in small $\theta_F$ with $\chi \approx i \alpha$ and then substitute the solution into Eq.\ (\ref{Ftheta}). This yields the following
expressions for the conductance and Fano factor:
\begin{gather}
 \frac{G}{G_0}
  = \mathcal{F}(0) = \frac{\xi \alpha}{2L} \coth \frac{\alpha}{2}, \label{QA_G} \\
 F
  = 1 - \left. \frac{2\mathcal{F}''(\theta)}{\mathcal{F}(\theta)} \right|_{\theta = 0}
  = \frac{\sinh\alpha - \alpha}{\sinh\alpha(\cosh\alpha - 1)}. \label{QA_F}
\end{gather}
They are plotted in Fig.\ \ref{ConductanceFanoFactor}. Expressions (\ref{QA_G}) and (\ref{QA_F}) were recently obtained in Ref.\ \onlinecite{Altland15} in the 
context of magnetotransport in Weyl semimetals when scattering between the Weyl nodes is neglected.

In the limit of balanced edge $\alpha = 0$, we recover the known values\cite{Beenakker92} of conductance $G/G_0 = \xi/L$ and Fano factor $F = 1/3$ for a
diffusive wire. In the limit of large $\alpha$, topologically protected channels dominate the transport giving the same value of $|\alpha|/2$ to all moments.
The Fano factor vanishes in this limit while the conductance equals $|m|/2$. The total conductance includes the contribution of the outer edges, Eq.\
(\ref{Gtotal}), and attains the value $G_\text{tot}/G_0 = \max\{n_L, n_R\}$ in the limit $\alpha \gg 1$.

Diffusion in the unprotected channels is exponentially suppressed at $\alpha \gg 1$. Remarkably, this non-perturbative localization-like effect occurs at
scales parametrically shorter than the actual localization length, $\xi/m \ll L \ll \xi$, and is accessible within the simple semiclassical treatment of the
sigma model.

\subsection{Mesoscopic conductance fluctuations}

Conductance of a disordered sample exhibits universal reproducible random fluctuations as a function of some external parameters, e.g., weak magnetic field or
gate voltage.\cite{Lee85,Lee87} These fluctuations are attributed to changes of the effective disorder potential and can be computed as the variation of 
conductance around its average value in the given ensemble (\ref{TQ_varG}). This quantity contains information about correlations of transmission probabilities 
of different channels \cite{Campagnano06, Nazarov09} and cannot be expressed in terms of the generating function $\mathcal{F}$ only.

In the quasiclassical limit $L \ll \xi$, the partition function factorizes: $Z = Z_B Z_F$. This approximation is insufficient to compute conductance 
fluctuations, since the parameters $\theta_B$ and $\theta_F$ are fully decoupled on the level of the minimized action. A more accurate calculation taking into 
account small fluctuations around the optimal trajectory is required. This will also involve Grassmann degrees of freedom of the sigma model.

In order to expand the action in small fluctuations, we parametrize the classical solution as
\begin{equation}
\label{Qc}
 Q_c = T_c^{-1} \Lambda T_c, \qquad \dot T_c T_c^{-1} = M = \text{const}.
\end{equation}
This representation is possible since the trajectory is a rotation with constant velocity around a suitably chosen axis, [cf.\ Eq.\ (\ref{smallerarc}) and Fig.\ 
\ref{Trajectory}]. Deviations from the optimal trajectory are encoded in the matrix $W$, which anticommutes with $\Lambda$:
\begin{equation}
 T = e^{W/2}\, T_c, \qquad Q = T_c^{-1} \Lambda e^W T_c. \label{W}
\end{equation}
We substitute these matrices into the action (\ref{NLSM_S}) and expand to the second order in $W$. This expansion yields $S = S_0 + S_1 + S_2$ with
\begin{subequations}
\label{S012}
\begin{align}
 S_0 &= S_\text{min}(\theta_F) - S_\text{min}(i\theta_B), \label{S0} \\
 S_1 &= \frac{\xi}{2L} \int_0^1 dx\, \str \Bigl[ (\Lambda M)^2 + \alpha \Lambda M \Bigr] W, \label{S1} \\
 S_2 &= \frac{\xi}{8L} \int_0^1 dx \str \Bigl[ \dot{W}^2 + 4 M \dot W W - \{\Lambda M, W\}^2 \nonumber\\
         &\qquad\qquad\qquad\qquad+\alpha \bigl( \Lambda \dot W W - 2 \Lambda M W^2 \bigr)  \Bigr]. \label{S2}
\end{align}
\end{subequations}
The linear term $S_1$ vanishes for the classical solution while $S_0$ yields the minimized action with $S_\text{min}$ defined by Eq.\ (\ref{QA_Smin}).

\begin{figure}
\center
\includegraphics[width=\columnwidth]{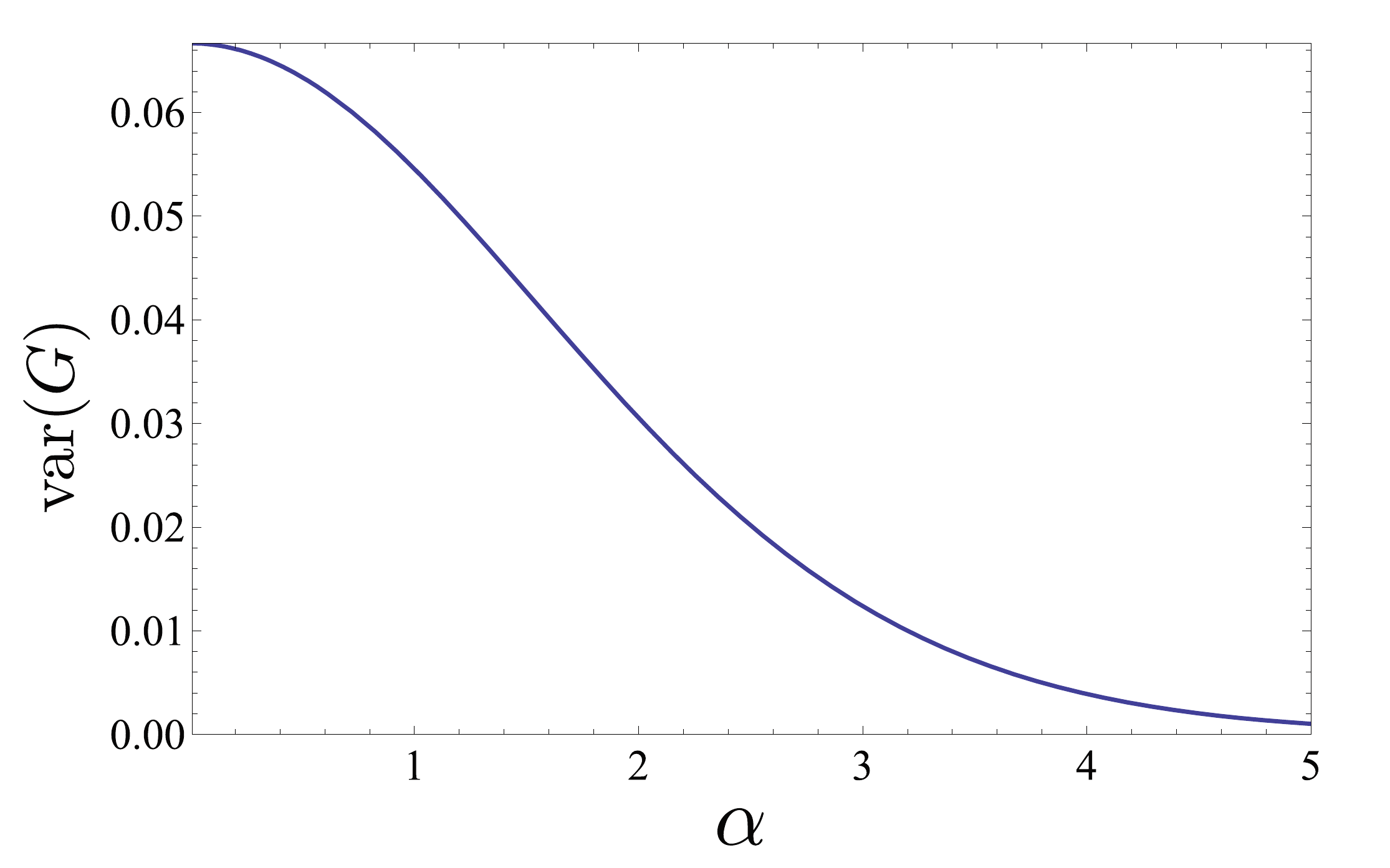}
\caption{Variance of conductance, Eq.\ (\ref{QA_UCF}), due to mesoscopic fluctuations as a function of $\alpha$.}
\label{varG}
\end{figure}

Further details on the parametrization of $W$ and computation of the fluctuations determinant are relegated to Appendix \ref{FM}. We should note that the
quadratic form in the components of $W$, corresponding to $S_2$, is difficult to diagonalize analytically for arbitrary values of $\theta_{B,F}$. However, in 
order to compute the variance of conductance [Eq.\ (\ref{TQ_varG})], an expansion in small $\theta_{B,F}$ suffices. This is possible within a perturbative 
treatment of the fluctuations determinant and yields
\begin{equation}
\label{QA_UCF}
\frac{\var G}{G_0^2} = \frac{\alpha^2(2 + \cosh\alpha) - 3\alpha \sinh \alpha}{16 \sinh^6(\alpha/2)}.
\end{equation}
Dependence of $\var G$ on $\alpha$ is shown in Fig.\ \ref{varG}. In the limit $\alpha = 0$, the known universal value $1/15$ is reproduced.\cite{Lee85, Lee87, 
Beenakker97, Mirlin94} For a spin degenerate sample, the variance is four times larger. In the presence of imbalance, $\alpha \neq 0$, some channels are 
topologically protected and have perfect transmission. Hence, the variance of conductance decreases with growing $\alpha$.

\subsection{Other symmetry classes}

Results for other symmetry classes with $\mathbb{Z}$ topology (classes C and D) are very similar to the results for the unitary class. The distribution
function, conductance, and Fano factor are given by exactly the same expressions (\ref{QA_rhol}), (\ref{QA_G}), and (\ref{QA_F}), respectively. The variance of
conductance also has the form (\ref{QA_UCF}) with an additional factor $2$ in classes C and D to account for particle-hole symmetry.

For the classes AII and DIII, the $\mathbbm{Z}_2$ topology is not captured on the quasiclassical level. The topological term in the action (\ref{NLSM_SAII}) is
quantized and thus drops from the classical equations of motion. Nevertheless, for fixed boundary conditions there are always two topologically distinct
trajectories minimizing the action in each of the two homotopy classes, (cf.\ Fig.\ \ref{SymplecticManifold}). The presence of the topological term becomes
crucial when both trajectories have approximately equal action. This happens when the angle $\theta_F$ is close to $\pi$. Hence, we conclude, that
$\mathbbm{Z}_2$ topology has an effect on the distribution function of transmission probabilities $\rho(\lambda)$ for small values of $\lambda$ (close to
perfect transmission). This limit will be discussed in the next section.

\section{Transmission distribution near $\lambda = 0$: mapping to random matrices}
\label{M0D}

The quasiclassical consideration of the previous section is valid provided the action has a well-defined minimum given by the solution of the classical 
equations
of motion. This is, however, not true when the final point of the trajectory is close to the ``south pole'' (the point with $\theta = \pi$). Hence, our result
for the distribution function (\ref{QA_rhol}) should be refined for small values of $\lambda$. Namely, we will demonstrate that the semiclassical gap in the 
distribution function is not exact and will obtain a more accurate result in the gap region.

\subsection{Illustrative example: A sphere}

Consideration of the distribution function at small $\lambda$ is similar for all symmetry classes. We start with the example of a particle moving on a sphere
and then generalize the result to a general sigma-model manifold.

\begin{figure}
\center
\includegraphics[width=0.2\textwidth]{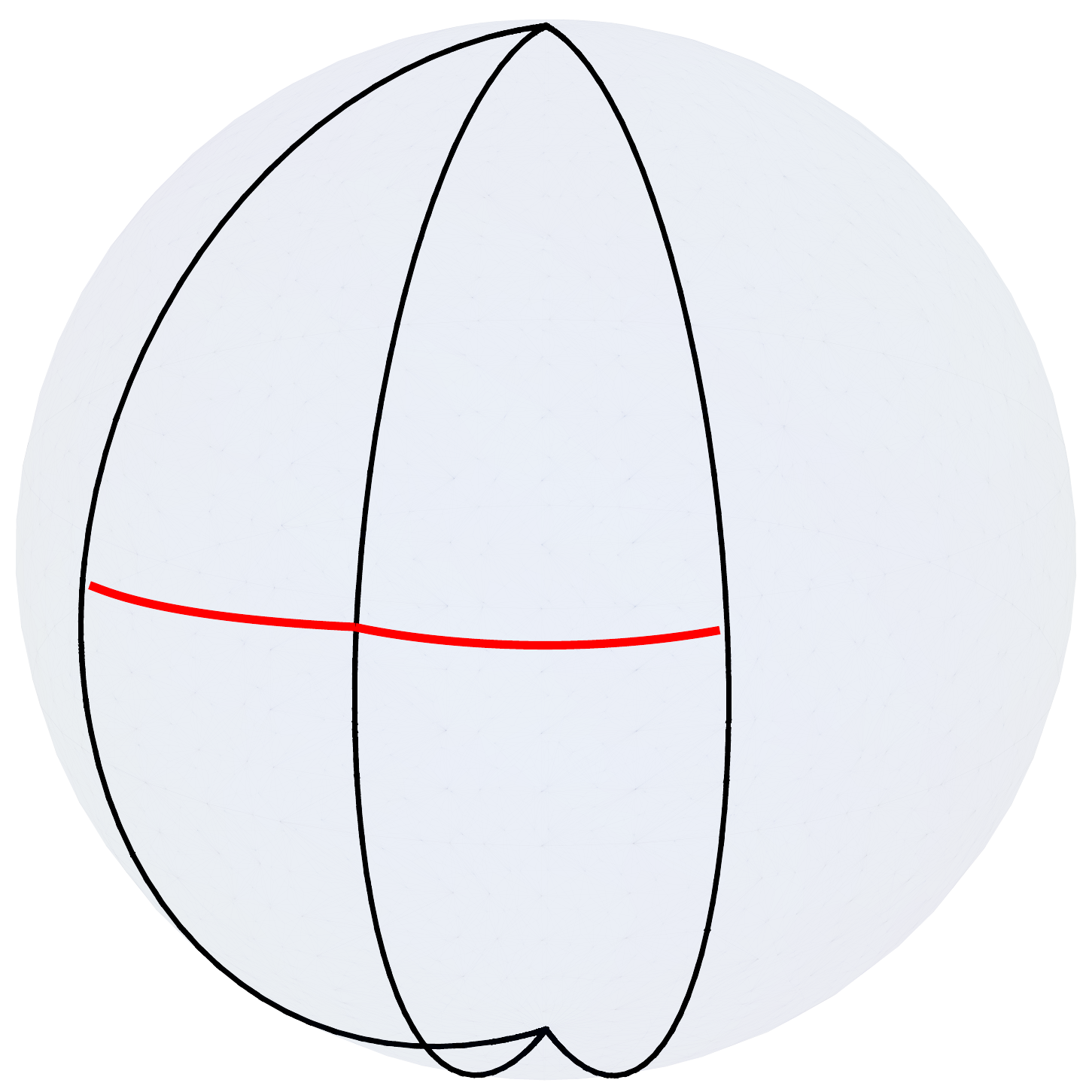}
\caption{Degenerate classical trajectories connecting the ``north pole'' to the ``south pole'' on a sphere. The soft mode corresponds to the azimuthal angle. 
Each trajectory is identified with a point on the ``equator''.}
\label{Deg}
\end{figure}

Consider the mechanical problem of a particle going from the ``north pole'' $\theta = 0$ to the ``south pole'' $\theta = \pi$ on a sphere. There are many
equivalent solutions to this problem, since all the ``meridians'' have exactly the same length (Fig.\ \ref{Deg}). This implies that the minimum of the classical
action is degenerate with respect to the azimuthal angle. If the final point of the trajectory is close to but not exactly at the ``south pole'',
the exact degeneracy is lifted.
However, there is still a soft mode approximately corresponding to the azimuthal angle.

Let us assume that the final point of the trajectory is at $\theta = \pi - \omega$ and $\phi = 0$. There are two classical solutions yielding a minimum and a
maximum value of the action: $\theta = (\pi \mp \omega) x$ with $\phi = 0$ or $\pi$. The two extremal values of the action are close provided $\omega \ll 1$.
We can interpolate between the two solutions by a family of trajectories parametrized by the angle $\phi$, that labels the point where the trajectory crosses
the ``equator'' of the sphere. Using Eq.\ (\ref{QA_S}) and assuming $\alpha$ small, we write effective action for this soft mode as
\begin{equation}
 \label{Ssphere}
 S
  = \text{const} - \frac{\pi \xi \omega}{2 L}\, \cos\phi + i m \phi.
\end{equation}
The last term appears due to $\alpha$ and is the relative magnetic flux enclosed by the trajectory.

The expansion (\ref{Ssphere}) is valid provided $\omega \ll \sqrt{L/\xi}$, when higher terms can be safely neglected in the weight function $e^{-S}$. The
partition function corresponding to the action (\ref{Ssphere}) has the form
\begin{equation}
Z(\omega) = \int_0^{2\pi} \frac{d \phi}{2\pi} e^{(\pi \xi \omega/2 L) \cos \phi - i m \phi} = I_m \left( \frac{\pi \xi \omega}{2 L}\right),
\end{equation}
where $I_m$ is the modified Bessel function. Thus we have effectively mapped the classical problem in one dimension to an effective zero-dimensional (0D) quantum problem.

\subsection{Generalization to symmetric superspaces}

We will now generalize the mapping discussed above to the case of a symmetric superspace. Consider first the case of a compact symmetric space. There is a
family of classical trajectories connecting the ``north pole'' $Q = \Lambda$ to the ``south pole'' $Q = -\Lambda$. Let us pick one particular geodesic
connecting the two poles (we again rescale $x$ by the length of the sample $L$),
\begin{equation}
\label{T0}
 Q = T_0^{-1} \Lambda T_0, \qquad T_0 = \exp(i \pi M_0 x/2).
\end{equation}
The matrix $M_0$ represents a point on the ``equator''. Other possible trajectories are generated by rotations from the small group $K$ of the matrices that
commute with $\Lambda$ and leave the end points $\pm\Lambda$ invariant.

We can always choose the generator $M_0$ such that it anticommutes with $\Lambda$. Since the trajectory (\ref{T0}) ends at the point $-\Lambda$, we have
$e^{i\pi M_0} = -1$ and conclude that eigenvalues of $M_0$ are $\pm 1$. (Other odd integer eigenvalues correspond to longer trajectories and hence yield larger
value of the action). The whole ``equator'' is parametrized by the matrix $M$ defined as
\begin{subequations}
\label{M}
\begin{gather}
 M = K^{-1} M_0 K, \qquad M^2 = 1, \label{M0} \\
 \{M, \Lambda\} = 0, \qquad [K,\Lambda] = 0.
\end{gather}
\end{subequations}
Hence, the ``equator'' of a symmetric space is also a symmetric space of a different class. ``Equators'' for all 10 symmetry classes are listed in Table 
\ref{Mapping}.

\begin{table}
\center
\begin{ruledtabular}
\begin{tabular}{ccccc}
$H$ & ``E'' & $Q_{FF}$ & $d$=1 & $d$=2\\
\hline
A & AIII & $\mathrm{U}(2n)/\mathrm{U}(n) \times \mathrm{U}(n)$ & 0 & $\mathbb{Z}$ \\
AIII& A & $\mathrm{U}(n) \times \mathrm{U}(n)/\mathrm{U}(n)$ & $\mathbb{Z}$ & 0 \\
\hline
AI& CI & $\mathrm{Sp}(4n)/\mathrm{Sp}(2n) \times \mathrm{Sp}(2n)$ & 0 & 0 \\
BDI& AI & $\mathrm{U}(2n)/\mathrm{Sp}(2n)$ & $\mathbb{Z}$ & 0 \\
D& BDI & $\mathrm{O}(2n)/\mathrm{U}(n)$ & $\mathbb{Z}_2$ & $\mathbb{Z}$ \\
DIII & D & $\mathrm{O}(n) \times \mathrm{O}(n)/\mathrm{O}(n)$ & $\mathbb{Z}_2$ & $\mathbb{Z}_2$\\
AII & DIII & $\mathrm{O}(2n)/\mathrm{O}(n) \times \mathrm{O}(n)$ & 0 & $\mathbb{Z}_2$ \\
CII & AII & $\mathrm{U}(n)/\mathrm{O}(n)$ & $\mathbb{Z}$ & 0\\
C & CII & $\mathrm{Sp}(2n)/\mathrm{U}(n)$ & 0 & $\mathbb{Z}$ \\
CI & C & $\mathrm{Sp}(2n) \times \mathrm{Sp}(2n)/\mathrm{Sp}(2n)$ & 0 & 0 \\
\end{tabular}
\end{ruledtabular}
\caption{Symmetry classification of the disordered sys\-tems.\cite{Altland97, Schnyder09, Kitaev09} Columns: Hamiltonian symmetry class, ``equator'' symmetry
class, compact part of the sigma-model manifold, possible topological insulators in 1D and 2D.}
\label{Mapping}
\end{table}

Generalization to a symmetric superspace is more subtle since the non-compact sector does not possess a ``south pole''. Nevertheless, calculation of 
transmission distribution function involves analytic continuation of the bosonic angle $\theta_B$ to the vicinity of $-i\pi$. This is equivalent to the ``south 
pole'' of the bosonic sector and means that the mapping from Table \ref{Mapping} applies in the same way to superspaces if a proper analytic continuation is 
assumed.

In order to derive an effective action for the trajectories connecting the ``north pole'' to the vicinity of the ``south pole'', we parametrize the $Q$ matrix
as
\begin{equation}
 \label{QW}
 Q = T^{-1} \Lambda (1 + W) T, \qquad T = \exp(i \pi M x/2).
\end{equation}
The matrix $W$ anticommutes with $\Lambda$ and describes a deviation of the trajectory from the ``meridian''. The final point of the trajectory is independent 
of $M$. We choose it at $Q(1) = -\Lambda (1 - i\hat\omega M_0)$ with the matrix $M_0$ from Eq.\ (\ref{M0}). The matrix $\hat\omega$ quantifies the deviation of 
the final point from the ``south pole'':
\begin{equation}
 \label{omega}
 \hat\omega
  = \begin{pmatrix}
      \pi - i\theta_B & 0 \\
      0 & \pi - \theta_F
    \end{pmatrix}_{\rm BF}.
\end{equation}
The value of $W$ at $x = 1$ satisfies
\begin{equation}
 \label{QF}
 W(1) = -i M \hat\omega M_0 M
\end{equation}
in order to ensure the correct final point of the trajectory.

We insert Eq.\ (\ref{QW}) into the action (\ref{NLSM_S}), neglect for the moment the topological term, and expand to linear order in
$W$. Using the properties (\ref{M}) and (\ref{QF}), we get the result
\begin{align}
 S &= -\frac{\xi}{8L} \int_0^1 dx\, \str \dot Q^2
   = \frac{i\pi\xi}{4L} \int_0^1 dx\, \str \bigl( M \dot W \bigr) \nonumber \\
   &= \frac{\pi \xi}{4L} \str \bigl( \hat\omega M_0 M \bigr).
\end{align}

In order to find the contribution of the topological term, we neglect $W$ and redefine the matrix $T$ in Eq.\ (\ref{QW}) such that at the final point $T(1)$
is independent of $M$. This can be achieved by a suitable $x$-dependent left rotation $R$ from the group $K$:
\begin{equation}
 T = R(x) \exp(i \pi M x/2), \quad R(0) = 1, \quad R(1) = M_0 M.
\end{equation}
With this definition, the expression for the topological term in Eq.\ (\ref{NLSM_S}) becomes
\begin{align}
 S_\text{top}
  &= -\frac{m}{2} \int_0^1 dx\, \str T^{-1} \Lambda \dot T
  = -\frac{m}{2} \int_0^1 dx\, \str \Lambda \dot R R^{-1} \nonumber \\
  &= -\frac{m}{2}  \str \bigl[ \Lambda \ln (M_0 M) \bigr].
\end{align}
Thus, we have successfully mapped the 1D sigma model, defined in terms of the matrix $Q$, to the 0D sigma model in terms of $M$:
\begin{equation}
 \label{mapped}
 S = \frac{1}{4\Delta} \str \bigl( \hat\omega M_0 M \bigr) - \frac{m}{2}  \str \bigl[ \Lambda \ln (M_0 M) \bigr].
\end{equation}

The model (\ref{mapped}) describes statistics of large random matrices with the average level spacing $\Delta = L/\pi\xi$. The correspondence between the 
manifolds of $Q$ and $M$ (the latter is the ``equator'' of the former) is detailed in Table \ref{Mapping}. This correspondence conforms to the Bott 
periodicity.\cite{Kitaev09} Whenever the 1D model contains a topological term, the corresponding 0D model also acquires a topological term of the same type.

\subsection{Distribution of transmission probabilities}

Let us now apply the correspondence derived above to the calculation of the transmission distribution $\rho(\lambda)$. We begin with the case of the imbalanced
quantum Hall edge, which belongs to the unitary symmetry class A. The matrix $\Lambda$ is defined in Eq.\ (\ref{Lambda}) and the matrices $M_0$ and $M$ we
represent as
\begin{equation}
 M_0 = \begin{pmatrix}  0 & 1 \\ 1 & 0 \end{pmatrix}_{\rm RA}, \qquad
 M = \begin{pmatrix}  0 & P \\ P^{-1} & 0 \end{pmatrix}_{\rm RA}.
\end{equation}
This choice satisfies all the constraints (\ref{M}). The unitary matrix $P$ belongs to the manifold of the sigma model in class AIII, (see Table \ref{Mapping}).
In terms of $P$, the 0D sigma-model action (\ref{mapped}) takes the form
\begin{equation}
 \label{S0D}
 S = \frac{1}{4\Delta} \str \bigl[ \hat\omega (P + P^{-1}) \bigr] + m  \str \ln P.
\end{equation}

Such a sigma model was studied before in the context of random chiral matrices with zero eigenvalues.\cite{Ivanov02, Verbaarschot93} We are interested in the
distribution function of transmission probabilities that is given by Eq.\ (\ref{rhotheta}):
\begin{equation}
 \label{rho0D}
 \rho(\lambda)
  = \frac{2}{\pi} \re \frac{\partial}{\partial \omega_F} \int DP\, e^{-S(P)} \Bigr|_{\omega_B = \omega_F = -2i\lambda + 0},
\end{equation}
where $\omega_{B,F}$ are the entries of $\hat\omega$, cf.\ Eq.\ (\ref{omega}). This distribution function exactly coincides with the spectral density of a 
random chiral matrix normalized to the average level spacing $\pi \Delta = L/\xi$.

For an illustration, we perform the calculation explicitly for class A.\cite{Verbaarschot93} The matrix $P$ in Eq.\
(\ref{S0D}) can be written as a product of a usual and a Grassmann matrix:
\begin{equation}
 P
  = \begin{pmatrix}
      e^a & 0 \\ 0 & e^{i b}
    \end{pmatrix} \begin{pmatrix}
      1 - \mu \nu & \nu \\ \mu & 1 + \mu \nu
    \end{pmatrix}.
\end{equation}
The measure on the superspace is especially simple in this parametrization:
\begin{equation}
 dP = \frac{da\, db\, d\mu\, d\nu}{4\pi}.
\end{equation}
We compute the partition function with the action (\ref{S0D}) and obtain the following expression:
\begin{multline}
 Z
  = \int dP\, e^{-S(P)} \\
  = \left( \omega_F \frac{\partial}{\partial \omega_F} - \omega_B \frac{\partial}{\partial \omega_B} \right)
    K_m\Bigl(\frac{\omega_B}{2\Delta}\Bigr) I_m\Bigl(\frac{\omega_F}{2\Delta}\Bigr).
\end{multline}
The distribution function is given by Eq.\ (\ref{rho0D}) and has the form\cite{Verbaarschot93}
\begin{subequations}
\label{Zclasses}
\begin{equation}
\label{M0D_rA}
\rho_{\text{A}}(u) = \frac{u}{2} \left[J_m^2(u) - J_{m+1}(u) J_{m-1}(u)\right] + |m| \delta(u),
\end{equation}
where we have rescaled the parameter by the level spacing: $u = \lambda/\Delta$, $\rho(u) = \Delta \rho(\lambda)$.

The result (\ref{M0D_rA}) is depicted in Fig.\ \ref{rA}. In the limit $\lambda \ll \sqrt{L/\xi}$, it refines the quasiclassical result (\ref{QA_rhol}).
Indeed, instead of identically vanishing $\rho(\lambda)$, suggested by the classical calculation inside the gap, the true asymptotics is $\rho \sim
\lambda^{2m+1}$. When the gap is not too large, $m \ll \sqrt{\xi/L}$, the function (\ref{M0D_rA}) also describes smoothly the crossover from the subgap
region, where $\rho$ is strongly suppressed, to the saturation $\rho \approx 1$ above the gap. This crossover occurs at $\lambda \approx m L/\pi\xi$, which is
equivalent to $\lambda_g \approx \alpha/\pi$ found in the semiclassical calculation (\ref{QA_lambdag}) for small $\alpha$. The oscillations in $\rho(\lambda)$,
emerging around $\lambda_g$, are caused by statistical repulsion between individual transmission probabilities, in full analogy with level repulsion in
random matrices.

\begin{figure}
\center
\begin{tabular}{c}
\includegraphics[width=0.95\columnwidth]{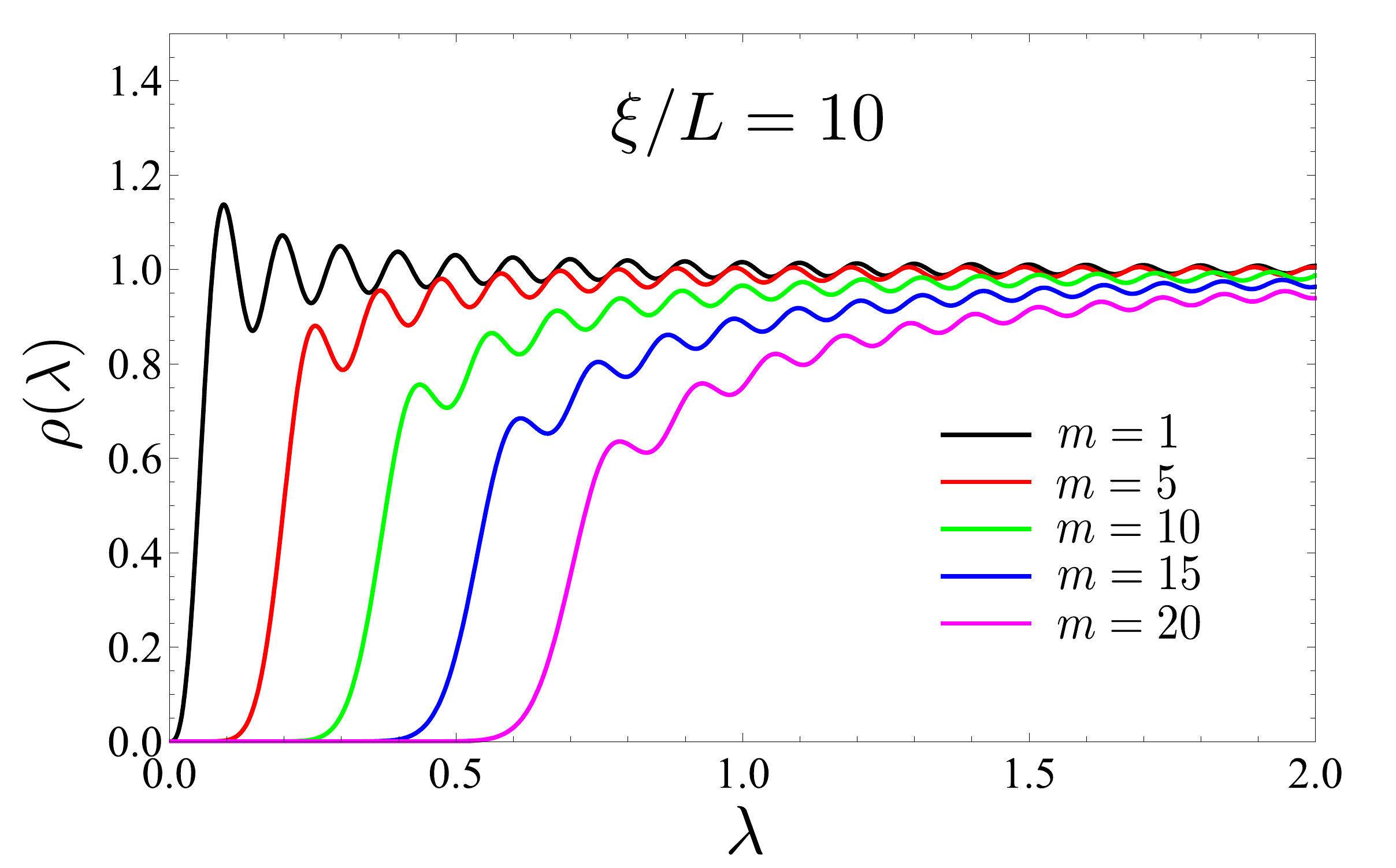} \\ \includegraphics[width=0.9\columnwidth]{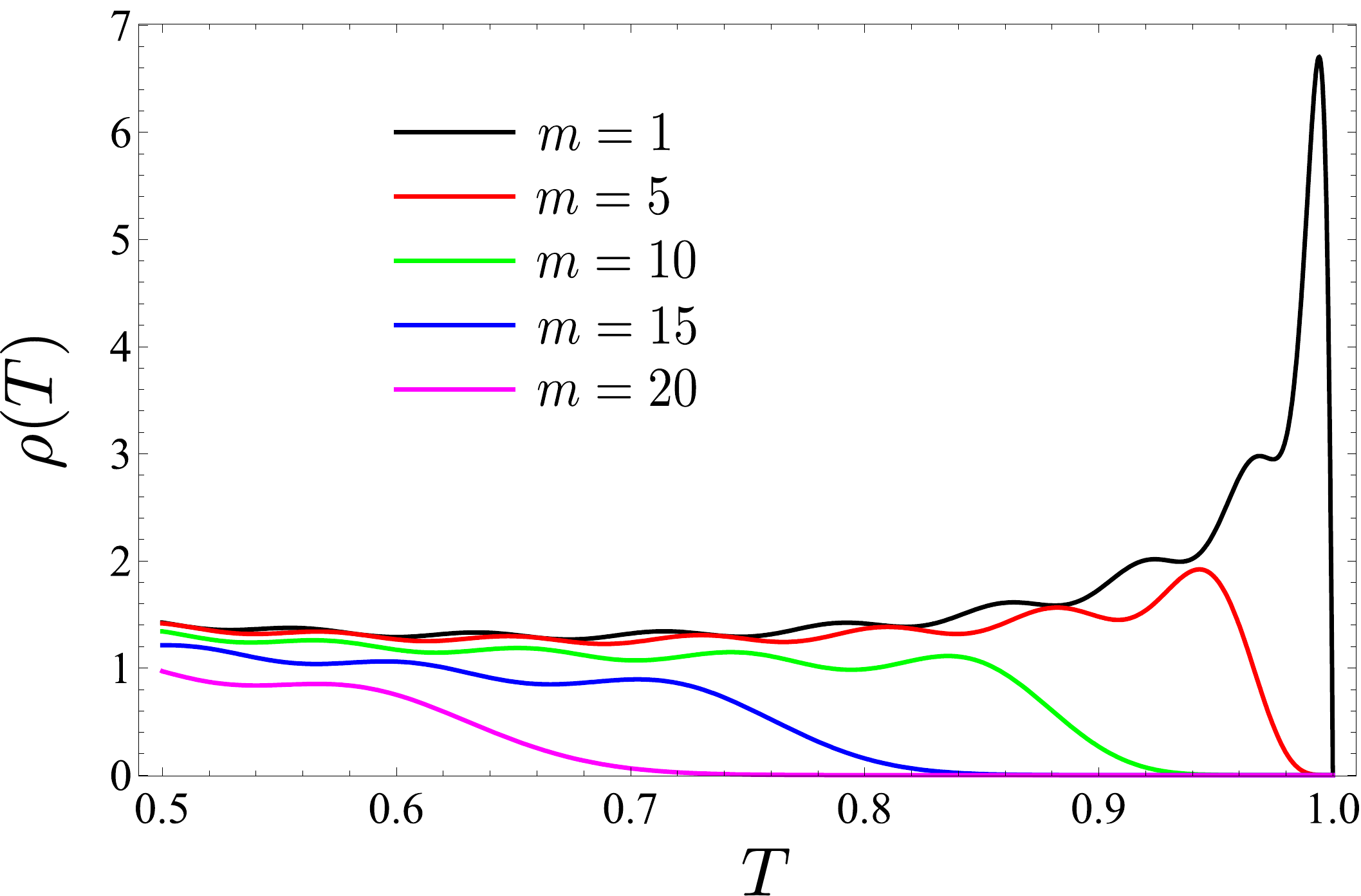}
\end{tabular}
\caption{Average distribution function of transmission probabilities in class A, [Eq.\ (\ref{M0D_rA})], at a fixed value of $\xi/L = 10$ and different $m$ as a function of $\lambda$ (upper panel) and $T$ (lower panel).}
\label{rA}
\end{figure}

\subsection{Other symmetry classes}

Exploiting the correspondence between the 1D transport problem and level statistics of random matrices, we can directly apply the results of Ref.\
\onlinecite{Ivanov02} to our problem. We introduce the normalized parameter $u = \lambda/\Delta$ and quote the results in terms of $\rho(u) = \Delta 
\rho(\lambda)$.
For classes C and D, we have
\begin{align}
\label{M0D_rC}
\rho_{\text{C}}(u) &= \rho_\text{A}(u) - \frac{1}{2}J_{m}(u) \int_0^u du'\, J_m(u'), \\
\label{M0D_rD}
\rho_{\text{D}}(u) &= \rho_\text{A}(u) + \frac{1}{2}J_m(u) \int_u^\infty du'\, J_m(u').
\end{align}
\end{subequations}
In these expressions, $m$ is any integer in class D and any even integer for class C.

\begin{figure}
\center
\begin{tabular}{c}
\includegraphics[width=0.93\columnwidth]{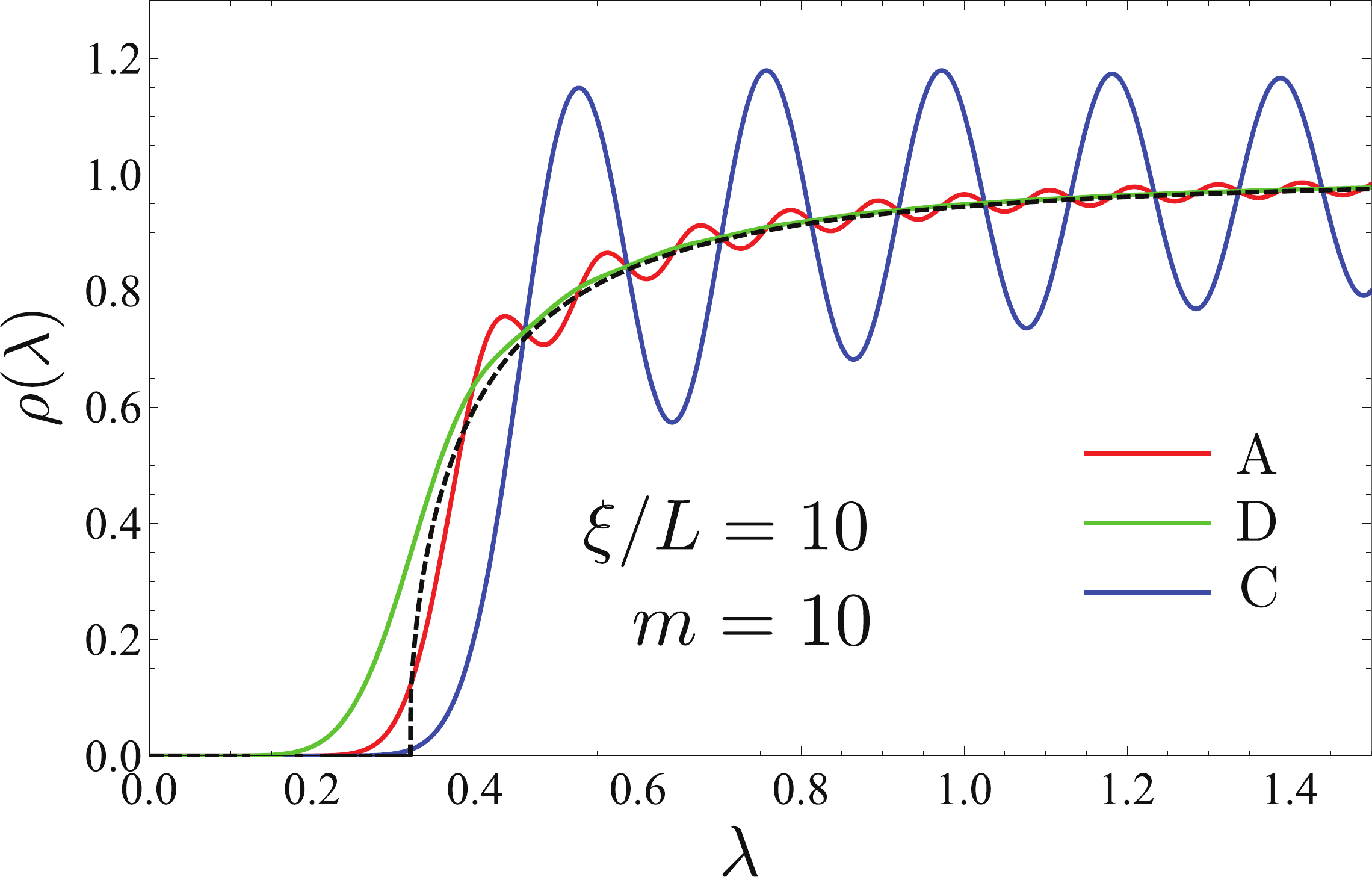} \\ \includegraphics[width=0.92\columnwidth]{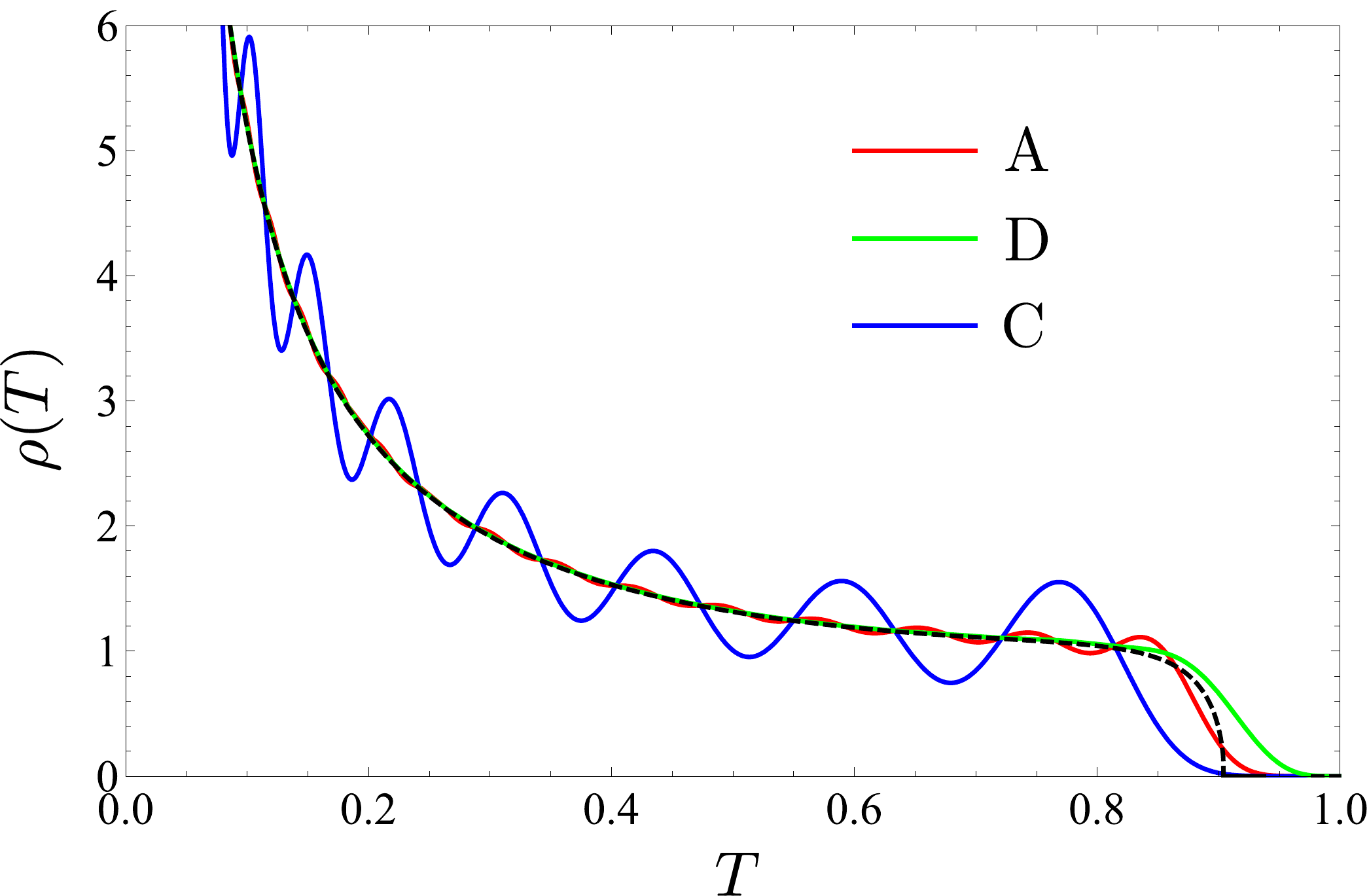}
\end{tabular}
\caption{Comparison of the average distribution function in classes A (quantum Hall edge), C, and D, [Eqs.\ (\ref{Zclasses})], with the semiclassical result (\ref{QA_rhol}) in terms of $\lambda$ (upper panel) and $T$ (lower panel).}
\label{rZ}
\end{figure}

All the above qualitative discussions of the results for the unitary class A apply to classes C and D as well. The only difference is in the strength of the
level repulsion. Classes C (spin quantum Hall effect) and D (thermal quantum Hall effect) exhibit stronger and weaker oscillations as compared to class A,
respectively. We compare the distribution functions (\ref{Zclasses}) and the semiclassical result (\ref{QA_rhol}) in Fig.\ \ref{rZ}.

The classes with $\mathbb{Z}_2$ topology, AII (quantum spin-Hall effect) and DIII, map to 0D models of the classes DIII and D, respectively. The latter are the
classes of random matrices with a possible single zero eigenvalue. The average spectral density for such matrices\cite{Ivanov02} yields the following result
for transmission distributions in 1D:
\begin{subequations}
\label{Z2classes}
\begin{align}
\label{M0D_rAII}
\rho_{\text{AII}}(u) &= \frac{u}{2} \left[J_1^2(u) + J_0(u) J'_1(u)\right] + \frac{\sigma}{2} J_1(u)\nonumber \\
& \qquad \qquad + (1 - \sigma) \delta(u), \\
\label{M0D_rDIII}
\rho_{\text{DIII}}(u) &= 1 + \sigma \frac{\sin u}{u} + (1 - \sigma) \delta(u).
\end{align}
\end{subequations}
The parameter $\sigma$ is either $1$ or $-1$ for the even and odd number of channels, respectively.

\begin{figure}
\center
\includegraphics[width=0.9\columnwidth]{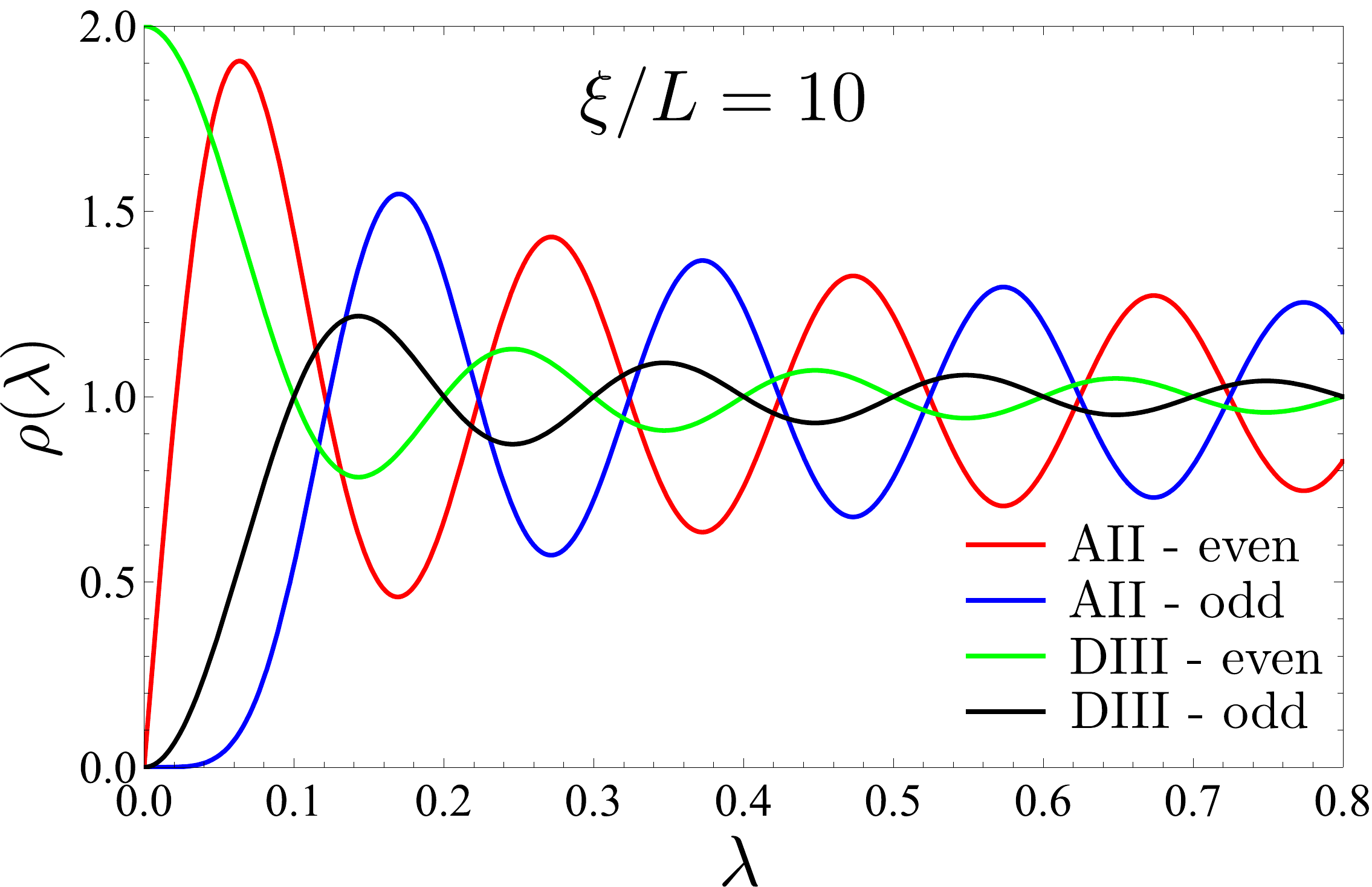}
\caption{Average distribution of transmission probabilities in classes AII (quantum spin-Hall edge) and DIII, Eqs.\ (\ref{Z2classes}), with even and odd number
of channels. In the odd case, a delta peak at $\lambda = 0$ appears.}
\label{rZ2}
\end{figure}

The results (\ref{Z2classes}) are depicted in Fig.\ \ref{rZ2}. A single channel with perfect transmission (delta function at $\lambda = 0$) emerges in the case 
of an odd number of channels. It suppresses $\rho(\lambda)$ at small $\lambda$, similarly to other classes discussed earlier. The mutual repulsion of 
transmission probabilities results in the oscillations in $\rho(\lambda)$ also in an analogy to the previously discussed classes. However, in the 
$\mathbb{Z}_2$ classes, the topological effects are weaker; this is the reason why they were not captured in the fully semiclassical analysis of Sec.\ \ref{QA}.

\subsection{Distribution function in the vicinity of the gap edge}
\label{BCG}

\begin{figure}
\center
\includegraphics[width=0.9\columnwidth]{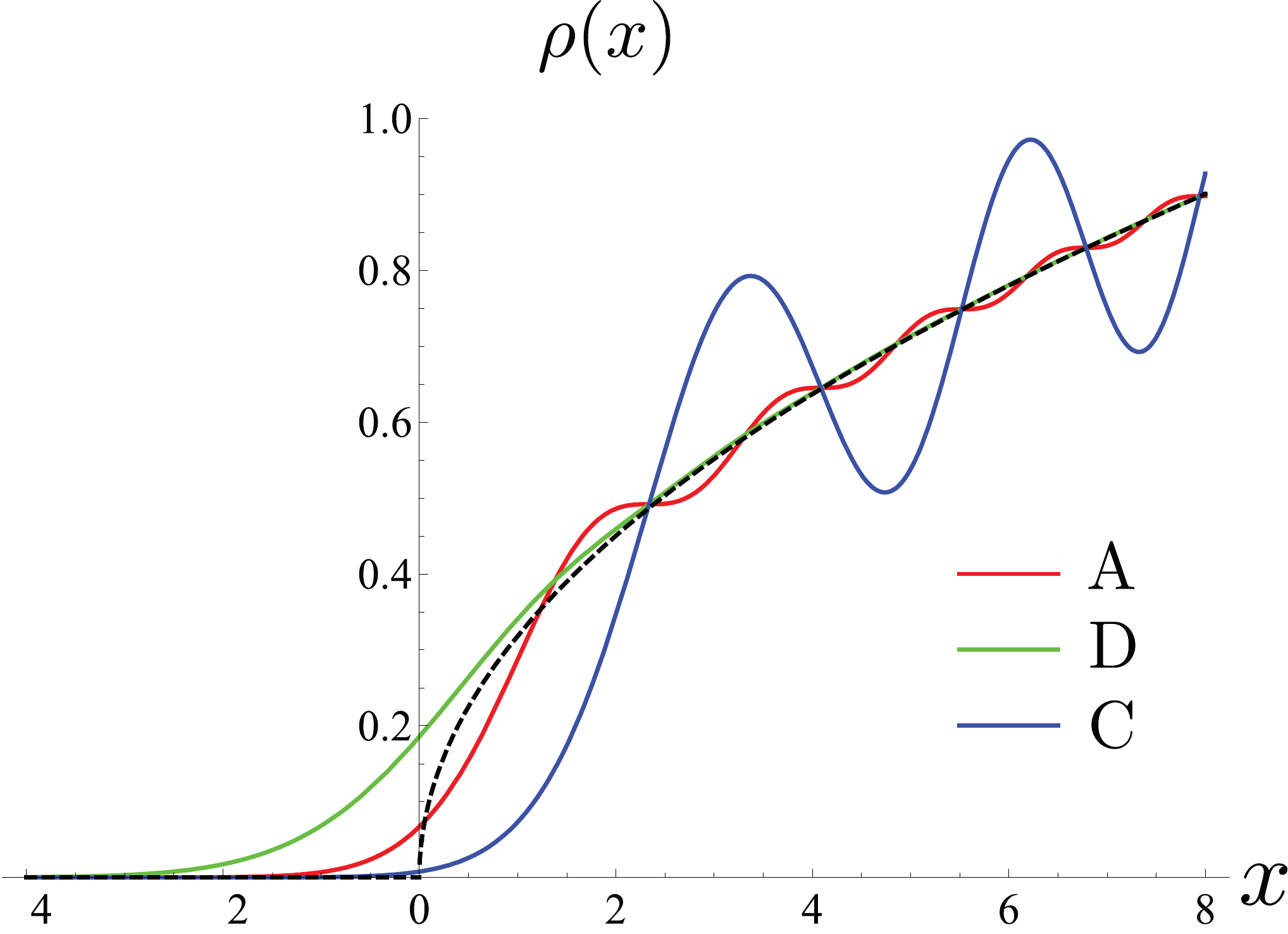}
\caption{Universal crossover functions close to the semiclassical edge of the spectrum (shown with dashed line) for classes A, C and D, Eqs.\
(\ref{TracyWidom}). The parameter $x$ is defined in (\ref{BCG_xl}).}
\label{rg}
\end{figure}

As was shown above, the behavior of the distribution function $\rho(\lambda)$ qualitatively changes from oscillatory to decaying as $\lambda$ is decreased
below $\lambda_g$. In order to describe this crossover quantitatively, let us introduce a suitably rescaled variable
\begin{equation}
 \label{x}
 x = \left(\frac{2}{m}\right)^{1/3} \left(\frac{\lambda}{\Delta} - m \right).
\end{equation}
In terms of this variable, we can extract the crossover dependence from Eqs.\ (\ref{Zclasses}) by invoking the asymptotic form of the Bessel function at $m
\gg 1$. This asymptotic expansion is given in terms of the Airy function:\cite{Nist10}
\begin{equation}
\label{M0D_Jmx}
J_m\left(m + x \left(\frac{m}{2}\right)^{1/3} \right) = \left(\frac{2}{m}\right)^{1/3} \! \! \Ai(-x) + O(1/m).
\end{equation}
The crossover functions for the three classes with $\mathbb{Z}$ topology have the form:
\begin{subequations}
\label{TracyWidom}
\begin{align}
\label{BCG_rgA}
\rho_{\text{A}}(x) &= x \Ai^2(-x) + [\Ai'(-x)]^2, \\
\label{BCG_rgC}
\rho_{\text{C}}(x) &= \rho_{\text{A}}(x) - \frac{1}{2} \Ai(-x) \int_{-\infty}^x \!\! \Ai(-t) \; dt,  \\
\label{BCG_rgD}
\rho_{\text{D}}(x) &= \rho_{\text{A}}(x) + \frac{1}{2} \Ai(-x) \int_x^\infty \!\! \Ai(-t) \; dt.
\end{align}
\end{subequations}
They are plotted in Fig.\ \ref{rg}.

These functions coincide with the spectral densities of the large-size random matrices close to the edge of the spectrum\cite{Tracy94, Deift07} in the three
standard Wigner-Dyson classes: orthogonal (AI), unitary (A), and symplectic (AII). Such a coincidence is not accidental. As long as we have mapped the 1D
transport problem onto the suitable random-matrix ensemble, the statistical properties in the vicinity of the spectral edge are universal.

At large positive $x$, all three functions (\ref{TracyWidom}) have the same square-root behavior
\begin{equation}
\label{BCG_rgs}
\rho(x) = \frac{\sqrt{x}}{\pi},
\end{equation}
shown in Fig.\ \ref{rg} by the dashed line. This envelope represents the edge of the distribution gap in the semiclassical solution (\ref{QA_rhol}).

It appears that the crossover functions (\ref{TracyWidom}) correctly describe the behavior of $\rho(\lambda)$ near the gap edge even in the limit $m \gg 
\sqrt{\xi/L}$. In this limit, the critical value $\lambda_g$ is too large and the mapping of Table \ref{Mapping} is not applicable for $\lambda \simeq 
\lambda_g$. On the semiclassical level, the appearance of the gap can be attributed to the existence of an unreachable region around the ``south pole'' of the 
sigma-model manifold, as is explained in Sec.\ \ref{QA}. When quantum fluctuations are taken into account, we expect the possibility of tunneling into this 
forbidden region. For $\lambda > \lambda_g$, there are always two classical solutions that represent a minimum and a maximum of the action. Exactly at $\lambda 
= \lambda_g$, the two solutions merge. For smaller values of $\lambda$ we again have two close ``classical'' solutions that extend into the complex plane. 
Since the corresponding minimum and maximum of the action are close, there is a soft mode interpolating between them. This means that the problem can be again 
mapped onto an effective 0D model taking into account only this soft mode. In the limit $1 \ll m \ll \sqrt{\xi/L}$, this mapping is equivalent to the mapping 
of Table \ref{Mapping}, hence, the universal crossover function can be derived from Eqs.\ (\ref{Zclasses}).

The calculation within the effective 0D model, describing the gap edge in the limit $m \gg \sqrt{\xi/L}$, is presented in Appendix \ref{Inst}. It yields the 
same crossover functions (\ref{TracyWidom}) provided the definition of the variable $x$ [Eq.\ (\ref{x})] is modified:
\begin{equation}
 \label{BCG_xl}
 x = \left(\frac{\xi}{L}\, f(\alpha) \right)^{2/3} (\lambda-\lambda_g) ,
\end{equation}
where the function $f(\alpha)$ is introduced in Eq.\ (\ref{QA_falpha}). The modified variable (\ref{BCG_xl}) is chosen such that the
semiclassical result maintains the form (\ref{BCG_rgs}) in the vicinity of the gap edge. Thus the single adjustable parameter in the universal dependence
(\ref{TracyWidom}) is extracted from the perturbative analysis of the semiclassical equations of motion.

\section{Summary and discussion}
\label{Summary}

We have studied the transport properties at the edge of a generic disordered two-dimensional topological insulator allowing for the coexistence of
topologically protected and diffusive channels. Two qualitatively different patterns of the topological effects emerge in the cases of $\mathbb{Z}$ and
$\mathbb{Z}_2$ topological insulators.

The prototypical model of disorder-mixed edge states with  $\mathbb{Z}$ topology is given by the junction between two quantum Hall states with different
filling factors (class A) (Fig.\ \ref{QH}). In this case, the average distribution function of transmission probabilities exhibits a semiclassical gap[ Eq.\
(\ref{QA_rhol})] shown in Fig.\ \ref{SemiClassicalRho}. The magnitude of the gap is related to the imbalance parameter $\alpha$ as shown in Fig.\
\ref{GapBeta}. The gap is accompanied by a delta peak at $\lambda = 0$ due to the topologically protected states [Eq.\ (\ref{delta})]. Strong suppression of
transmission in the unprotected channels occurs already at relatively short scales $\xi/|m| \ll L \ll \xi$. This non-perturbative effect is thus fully
accessible within the semiclassical treatment of the non-linear sigma model.

An exemplary model of a $\mathbb{Z}_2$ topological insulator with $N \gg 1$ edge modes is provided by a relatively thick quantum spin-Hall sample (class AII) (see Fig.\ \ref{QSH}). In this case, at most one edge channel is topologically protected leading to weaker transport effects.  In order to capture
topologically-driven suppression of nearly perfect transparencies in the average distribution function $\rho(\lambda)$, we have developed a mapping of the 1D
non-linear sigma model onto an equivalent 0D random matrix theory. This mapping applies to all symmetry classes and is summarized in Table \ref{Mapping}.
Specifically, the average distribution function $\rho(\lambda)$ is equivalent to the average spectral density of a certain random matrix ensemble. The latter
is described by the 0D sigma model defined on the ``equator'' of the original 1D sigma-model manifold. This mapping yields detailed description of the
transmission eigenvalues statistics in the vicinity of $\lambda = 0$ both for $\mathbb{Z}$ [Eqs.\ (\ref{Zclasses}), Fig.\ \ref{rZ}] and $\mathbb{Z}_2$ [Eqs.\
(\ref{Z2classes}), Fig.\ \ref{rZ2}] topological insulators.

When the size of the sample exceeds the localization length, $L \gg \xi$, the average distribution $\rho(\lambda)$ qualitatively changes. Repulsion
between transmission probabilities gets strong and the oscillations of the type of Fig.\ \ref{rA} develop into sharp isolated peaks. This phenomenon is referred
to as ``crystallization'' of transmission eigenvalues.\cite{Frahm95,Lamacraft04,Altland05} This limit can be accessed within the non-linear sigma model 
(\ref{NLSM_S}) when it is studied
beyond the semiclassical approximation. This amounts to analyzing the full spectrum of the Laplace-Beltrami operator on the curved superspace.\cite{Mirlin94,
Rejaei96} The presence of topologically protected modes, and hence the topological term in the sigma-model action, should be also taken into account in this
case. This will be a subject of a separate publication.

The mapping from 1D sigma model to the equivalent 0D theory on the ``equator'' (Table \ref{Mapping}) can be extended to higher dimensions. In particular,
transport properties of the surface states of a 3D topological insulator can be described in terms of the effective 1D sigma-model with a topological and a
mass term. The latter encodes the information on source fields. Such an analysis requires the knowledge of the spectral properties of the corresponding
transfer-matrix Hamiltonian and will be discussed elsewhere.

\acknowledgments

We are grateful to I.\ Gornyi, A.\ Mirlin, E.\ K\"onig, and D.\ Polyakov for valuable discussions. The work was supported by Russian Science Foundation (Grant 
No.\ 14-42-00044).

\appendix

\section{Fluctuation modes}
\label{FM}

In this appendix, we compute the contribution of fluctuation modes around the classical solution of the sigma model (\ref{NLSM_S}) in the unitary symmetry
class. This calculation yields the result (\ref{QA_UCF}) for the mesoscopic conductance fluctuations.

In the presence of the topological term, the classical trajectory can be represented as a rotation with a constant speed around a certain axis that
depends on $\alpha$. More specifically, we can choose $T_c$, introduced in Eq.\ (\ref{Qc}), in the form
\begin{equation}
\label{FM_Tc}
T_c = \cos(\hat\chi x/2) + i \sin(\hat\chi x/2) (\tau_z \cos \hat{\psi} + \tau_y \sin \hat{\psi}).
\end{equation}
Here, $\tau_{x,y,z}$ are Pauli matrices in the RA space and $\hat{\chi}$ and $\hat\psi$ define the velocity and the axis of rotation, respectively. They are both
diagonal matrices in the BF space. Entries of $\hat\chi$ are solutions to Eq.\ (\ref{QA_chitheta}) for angles $i\theta_B$ and $\theta_F$, respectively. The
matrix $\hat\psi$ satisfies $\hat{\chi} \cos \hat{\psi} = i \alpha$ [cf.\ Eq.\ (\ref{smallerarc})]. We also introduce the matrix
\begin{equation}
\hat{\gamma} = \begin{pmatrix} \gamma_B & 0 \\ 0 & \gamma_F \end{pmatrix}_{\rm BF} = \hat{\chi} \sin \hat{\psi} = \sqrt{\hat{\chi}^2 + \alpha^2}.
\label{gamma}
\end{equation}
With these definitions, we write the generator of rotation\ (\ref{Qc}), as
\begin{equation}
M = \frac{i \hat{\chi}}{2} (\tau_z \cos \hat{\psi} + \tau_y \sin \hat{\psi}) = \frac{i \hat\gamma \tau_y - \alpha \tau_z}{2}.
\end{equation}
It can be checked by a direct calculation, that the linear action $S_1$ [Eq.\ (\ref{S1})] vanishes with such $M$.

As long as the matrix $W$ anticommutes with $\Lambda$, we parametrize $W = w_x \tau_x + w_y \tau_y$ and write the quadratic action $S_2$, Eq.\ (\ref{S2}), in 
the form
\begin{multline}
 S_2
  = \frac{\xi}{4L} \int_0^1 dx \str \Bigl[
      (\dot{w}_x^2 + \dot{w}_y^2) - i \alpha (\dot{w}_x w_y - \dot{w}_y w_x) \\
      -\frac{\hat{\gamma}^2}{2} (w_x^2 + w_y^2) - \frac{1}{2} [(\hat{\gamma} w_x)^2 - (\hat{\gamma} w_y)^2]
    \Bigr].
\end{multline}

Using explicit parametrization in the BF space
\begin{equation}
\label{FM_wxwy}
w_x = \begin{pmatrix} p & \sigma \\ \kappa & i q \end{pmatrix}_{\rm BF}, \qquad w_y = \begin{pmatrix} m & \mu \\ \eta & i n \end{pmatrix}_{\rm BF},
\end{equation}
we recast the quadratic action in the form
\begin{multline}
 S_2
  = \frac{\xi}{4L} \int_0^1 dx \Biggl[
      \Bigl( p\; m \Bigr) H_B  \begin{pmatrix} p \\ m \end{pmatrix}
      +\Bigl( q\; n \Bigr) H_F  \begin{pmatrix} q \\ n \end{pmatrix} \\
      +2 \Bigl( \sigma\; \mu \Bigr) H_{\rm BF}  \begin{pmatrix} \kappa \\ \eta \end{pmatrix}
    \Biggr]
 \label{FM_S2}
\end{multline}
with the operators
\begin{subequations}
\label{FM_H}
\begin{gather}
H_{B,F} = \begin{pmatrix}
         -\dfrac{\partial^2}{\partial x^2} - \gamma_{B,F}^2 & i\alpha \dfrac{\partial}{\partial x} \\[10pt]
         -i\alpha \dfrac{\partial}{\partial x} & -\dfrac{\partial^2}{\partial x^2}
       \end{pmatrix}, \label{FM_SBSF} \\
H_{\rm BF} = \begin{pmatrix}
         -\dfrac{\partial^2}{\partial x^2} - \dfrac{(\gamma_B + \gamma_F)^2}{4} & i\alpha \dfrac{\partial}{\partial x} \\[10pt]
         -i\alpha \dfrac{\partial}{\partial x} & -\dfrac{\partial^2}{\partial x^2} - \dfrac{(\gamma_B - \gamma_F)^2}{4}
       \end{pmatrix}. \label{FM_SBF}
\end{gather}
\end{subequations}

The partition function of the sigma model can be written in terms of corresponding functional determinants as
\begin{equation}
 Z
  = \frac{\det H_{\rm BF}}{\sqrt{\det H_B \det H_F}}\; e^{-S_0}.
\end{equation}
This function obeys the supersymmetry condition since at $\gamma_B = \gamma_F$ we have $H_B = H_F = H_{\rm BF}$ and hence $Z = 1$.

In order to calculate the variance of conductance, Eq.\ (\ref{TQ_varG}), we will take the mixed derivative in both $\theta_B$ and $\theta_F$. Hence, only the
Grassmann determinant $H_{BF}$ contributes
\begin{equation}
 \frac{\var G}{G_0^2} = 4\, \frac{\partial^4 \ln\det H_{BF}}{\partial^2 \theta_B\; \partial^2 \theta_F}  \Big|_{\theta_F=\theta_B=0}.
\end{equation}
Diagonalizing the operator $H_{BF}$ under the boundary conditions $W(0) = W(1) = 0$ is a tedious problem. However, for our purposes it is sufficient to carry
out the calculation for small values of $\theta_{B,F}$. Approximately solving Eq.\ (\ref{QA_chitheta}), we obtain the relation between $\gamma$ and $\theta$
valid at small angles,
\begin{equation}
 \gamma_{B,F} = \frac{\alpha\, \theta_{B,F}}{2\sinh(\alpha/2)}.
\end{equation}

To facilitate the calculation, we perform the gauge transformation $\tilde{H} = e^{\alpha x \sigma_y/2} H_{BF} e^{-\alpha x \sigma_y/2}$ that removes 
derivatives from the off-diagonal elements in Eq.\ (\ref{FM_SBF}). Expanding the determinant in small $\gamma_{B,F}$, we arrive at the following expression for 
the variance of conductance:
\begin{equation}
\label{FM_varg}
\frac{\var G}{G_0^2} = \frac{\alpha^4}{8 \sinh^4(\alpha/2)} \tr \Bigl[ H_0^{-2} + 2 H_0^{-1} e^{\alpha x} H_0^{-1} e^{-\alpha x} \Bigr].
\end{equation}
Here the Hamiltonian is $H_0 = -\partial^2/\partial x^2 + \alpha^2/4$, and we can evaluate the traces by expanding in the eigenfunctions $\sqrt{2} \sin (\pi l
x)$ for integer $l \geq 1$. This yields
\begin{gather}
\tr H_0^{-2} = \frac{1}{\alpha^4} \left[-8 + 2\alpha \coth \frac{\alpha}{2} + \frac{\alpha^2}{\sinh^2(\alpha/2)} \right], \\
\begin{multlined}
 \tr\bigl[H_0^{-1} e^{\alpha x} H_0^{-1} e^{-\alpha x}\bigr] \\
   = \frac{1}{2 \alpha^4} \left[8 -5 \alpha \coth \frac{\alpha}{2} + \alpha^2 + \frac{\alpha^2}{2 \sinh^2(\alpha/2)} \right].
\end{multlined}
\end{gather}
Substituting into Eq.\ (\ref{FM_varg}), we obtain the result (\ref{QA_UCF}) for the conductance variance.

\section{Fluctuations near $\lambda = \lambda_g$}
\label{Inst}

In this Appendix, we consider the contribution of soft fluctuation modes to the distribution function $\rho(\lambda)$ near the semiclassical threshold value 
$\lambda = \lambda_g$ [cf.\ Eq.\ (\ref{QA_rholg})].

The size of the semiclassical gap $\lambda_g$ is determined by Eq.\ (\ref{QA_chitheta}) with $\theta_F = \pi + 2 i \lambda_g$ and $\chi_g$ is given by Eq.\ 
(\ref{QA_gap}) as a function of $\alpha$. At the critical value $\lambda = \lambda_g$, two classical solutions (minimum and maximum of the action) merge, hence,
a small deviation $\Delta\lambda = \lambda - \lambda_g$ scales as the square of $\Delta\chi = \chi - \chi_g$. Expanding Eq.\ (\ref{QA_chitheta}) near 
$\lambda_g$ and $\chi_g$ and using Eq.\ (\ref{QA_gap}), we derive the relation
\begin{equation}
 \Delta\chi
  = \frac{i}{\chi_g} \sqrt{\frac{\Delta\lambda + i0}{A B}},
 \label{chilambda}
\end{equation}
where we have introduced the following two constants:
\begin{equation}
 A = \sqrt{\frac{\chi_g^2 \sin\chi_g}{\chi_g - \sin\chi_g}},
 \quad
 B = \frac{2\chi_g + \chi_g \cos\chi_g - 3\sin\chi_g}{4 \chi_g^4 \sin\chi_g}.
 \label{AB}
\end{equation}
Such a separation of the factors will be convenient in the subsequent calculations.

We expand the minimized classical action on the sphere, Eq.\ (\ref{QA_Smin}), in powers of $\Delta\lambda$. The first derivative of the action $\partial 
S_\text{min}/\partial \theta_F$ provides the value of the generating function $\mathcal{F}(\theta)$ in Eq.\ (\ref{Ftheta}). The next expansion term can be 
derived assuming a small deviation $\Delta\chi$ in Eq.\ (\ref{Ftheta}). Using relation (\ref{chilambda}), we obtain
\begin{align}
 S_\text{min}
  &= \frac{\xi}{L}\left[
      \mathrm{const} - A \Delta\lambda + \frac{2 \chi_g}{3 A} \Delta\lambda \Delta\chi
    \right] \nonumber \\
  &= \frac{\xi}{L}\left[
      \mathrm{const} - A \Delta\lambda + \frac{2 i}{3 \sqrt{B}} \left(\frac{\Delta\lambda}{A}\right)^{3/2}
    \right].
 \label{Sminlambda}
\end{align}
This action correctly reproduces semiclassical square-root behavior of the density (\ref{QA_rholg}) with the prefactor $f(\alpha) = A^{-3/2} B^{-1/2}$ [cf.\ Eq.\ (\ref{QA_falpha})]. The full supersymmetric minimized action is given by Eq.\ (\ref{S0}):
\begin{equation}
 S_0
  = \frac{\xi}{L} \str \left[
      A \Delta\hat\lambda - \frac{2 \chi_g}{3 A} \Delta\hat\lambda \Delta\hat\chi
    \right].
 \label{A_S0}
\end{equation}
Here, we use the notation $\hat\lambda = \diag\{\lambda_B, \lambda_F\}$ and similar for $\hat\chi$.

Let us consider small fluctuations around the classical minimum of the action. These fluctuations are described by the matrix $W$ [Eq.\ (\ref{W})], and the 
expansion of the action up to the second order in $W$ is given by Eqs.\ (\ref{S012}). The linear term $S_1$ vanishes at the minimum of the action. The quadratic 
term $S_2$ can be written explicitly in components of $W$ as in Eq.\ (\ref{FM_S2}) with the Hamiltonian (\ref{FM_H}) and parameters $\gamma_{B,F}$ defined by 
Eq.\ (\ref{gamma}). Exactly at the edge of the classical gap, $\gamma_B = \gamma_F = \gamma_g$, all three Hamiltonians (\ref{FM_H}) coincide and possess a zero 
mode which signals that two classical solutions merge at $\lambda_g$. The zero mode has the following two-component wave function:
\begin{align}
\label{Inst_uv}
 u(x)
  &= \cos\left(\chi_g x - \frac{\chi_g}{2}\right) - \cos\frac{\chi_g}{2}, \\
 v(x)
  &= \frac{i \gamma_g^2}{\alpha}\, x \cos\frac{\chi_g}{2} - \frac{i \alpha}{\chi_g} \left[
      \sin \left(\chi_g x - \frac{\chi_g}{2}\right) + \sin\frac{\chi_g}{2}
    \right]. \nonumber
\end{align}
It manifestly satisfies the conditions $u(0) = u(1) = v(0) = 0$ while the condition $v(1) = 0$ follows from the gap equation (\ref{QA_gap}).

With small deviations $\Delta\lambda_{B,F}$, the mode (\ref{Inst_uv}) acquires a small mass of the order of $\Delta\chi \sim \sqrt{\Delta\lambda}$. We will 
retain only this soft mode in the action and parametrize relevant fluctuations by
\begin{equation}
 W = \bigl[ u(x) \tau_x + v(x) \tau_y \bigr] c.
\end{equation}
Here, $c$ is a constant $2\times 2$ matrix in the BF space while the coordinate dependence is contained in the functions $u$ and $v$. Quadratic action can be 
now computed by assuming small perturbations $-2\gamma_g\Delta\gamma = -2\chi_g \Delta\chi$ in the operators (\ref{FM_H}). This yields
\begin{equation}
 S_2
  = \frac{\xi X}{L} \str (\chi_g \Delta\hat\chi c^2).
 \label{A_S2}
\end{equation}
with the factor
\begin{equation}
 X
  = -\frac{1}{2} \int_0^1 dx\, u^2(x)
  = -\chi_g^3 \sin\chi_g\; B.
\end{equation}
and $B$ from Eq.\ (\ref{AB}).

Since the quadratic term in the action (\ref{A_S2}) vanishes at $\lambda = \lambda_g$, we expand the action further taking into account the cubic term,
\begin{multline}
 S_3
  = \frac{\xi}{4L} \int_0^1 dx \str \biggl[
      M \{ W^2, \dot W \} + \Lambda M W \Lambda M W^2 \\
      +\frac{1}{3} (\Lambda M)^2 W^3 + \frac{\alpha}{3} \Lambda M W^3
    \biggr]
  = -\frac{\xi Y}{3 L} \str c^3.
 \label{A_S3}
\end{multline}
The factor $Y$ can be computed directly at $\lambda = \lambda_g$ and takes the value
\begin{equation}
 Y = -\frac{3\gamma_g}{4} \int_0^1 dx (u^2 + v^2)(2 i \dot v - \alpha u) = (\chi_g^3 \sin\chi_g)^{3/2} B.
\end{equation}

We have thus constructed the relevant expansion of the action both in $\Delta\lambda$ and in fluctuations $c$. Collecting together the terms (\ref{A_S0}), 
(\ref{A_S2}), and (\ref{A_S3}) we obtain
\begin{equation}
 S
  = \frac{\xi}{L} \str \left[
      A \Delta\hat\lambda - \frac{2 \chi_g}{3 A} \Delta\hat\lambda \Delta\hat\chi
      + X \chi_g \Delta\hat\chi c^2 - \frac{Y c^3}{3}
    \right].
 \label{Sc}
\end{equation}
This action can be simplified in terms of a rescaled matrix field $h$ and parameters $\hat x = \diag\{x_B, x_F\}$. The new variables are introduced as
\begin{equation}
 c = \frac{h (\xi B/L)^{-1/3} - \chi_g \Delta\hat\chi}{\sqrt{\chi_g^3 \sin\chi_g}},
 \quad
 \Delta\hat\lambda = A B^{1/3} (\xi/L)^{-2/3} \hat x.
 \label{hx}
\end{equation}
Note that the relation between $\hat x$ and $\hat \Delta\lambda$ coincides with Eq.\ (\ref{BCG_xl}) due to the identity $f(\alpha) = A^{-3/2} B^{-1/2}$ [cf.\ 
Eq.\ (\ref{QA_falpha})].

In the new variables (\ref{hx}), the action (\ref{Sc}) acquires a universal cubic form:
\begin{equation}
 S
  = \str\left[
      A^2 (\xi B/L)^{1/3} \hat x - \hat x h - \frac{h^3}{3}
    \right].
 \label{Sh}
\end{equation}
The matrix $h$ and the corresponding integration measure $dh$ can be written explicitly as
\begin{equation}
 h
  = \begin{pmatrix}
      p & \eta \\ \kappa & q
    \end{pmatrix}_{BF},
 \qquad
 dh
  = \frac{dp\, dq\, d\eta\, d\kappa}{2\pi i}.
\end{equation}

Integration contours for the variables $p$ and $q$ should be chosen in accordance with the structure of the original sigma-model manifold and deformed in order 
to ensure convergence of the integral with the action (\ref{Sh}). This leads to the following choice of the integration contours: $p:(-\infty,i\infty+0)$, 
$q:(-i\infty-0,i\infty-0)$. The partition function for the action (\ref{Sh}) can be now calculated in terms of the Airy functions
\begin{multline}
 Z(x_B, x_F)
  = \int dh\, e^{-S(h)}
  = \pi e^{A^2 (\xi B/L)^{1/3}(x_F - x_B)} \\
    \times \left( \frac{\partial}{\partial x_F} - \frac{\partial}{\partial x_B} \right) \Ai(-x_F)[\Bi(-x_B) - i \Ai(-x_B)].
 \label{ZAiry}
\end{multline}
This partition function obeys the supersymmetry relation $Z(x,x) = 1$.

The distribution of transmission probabilities follows from Eq.\ (\ref{rhotheta})
\begin{equation}
 \rho(x)
  = \frac{1}{\pi} \im \left. \frac{\partial Z}{\partial x_F} \right|_{x_B = x_F = x}.
\end{equation}
The exponential prefactor in the partition function (\ref{ZAiry}) is purely real and hence does not contribute to the distribution function. The resulting 
expression for $\rho(x)$ reproduces Eq.\ (\ref{BCG_rgA}).

\bibliographystyle{apsrev4-1}
\bibliography{refs}

\end{document}